\documentclass[sigconf]{acmart}

\usepackage{amsmath}
\usepackage{bm}
\usepackage{graphicx}
\usepackage{xcolor}
\usepackage{import}
\usepackage[utf8]{inputenc}
\usepackage{listings}
\usepackage[noend]{algpseudocode}
\usepackage{url}
\usepackage{multirow}
\usepackage[font={small,it}]{caption}
\usepackage{array}
\usepackage{rotating}   
\usepackage{makecell}
\usepackage{colortbl}
\usepackage{enumerate}
\usepackage{booktabs}
\usepackage{subfig}
\usepackage{paralist}
\usepackage[ruled,vlined]{algorithm2e}
\usepackage{amsfonts}
\usepackage{commath}
\usepackage{bm}
\usepackage{framed}

\algrenewcommand\algorithmiccomment[1]{\hfill \textcolor{gray}{$\triangleright$ \textit{#1}}}

\newenvironment{proof1} {\begin{proof}[Justification]} {\end{proof}}

\DeclareUnicodeCharacter{00A0}{ }

\newcommand{\varDash}[1]{{\operatorname{\mathit{#1}}}}

\newcommand{\specialcell}[2][c]{
	\begin{tabular}[#1]{@{}l@{}}#2\end{tabular}}

\newcommand*{\dittoclosing}{---''---}

\newcommand{\plusplus}[1]{\mathrel{{#1}{+}{+}}}

\makeatletter
\g@addto@macro\normalsize{
  \setlength\abovedisplayskip{4pt}
  \setlength\belowdisplayskip{4pt}
  \setlength\abovedisplayshortskip{4pt}
  \setlength\belowdisplayshortskip{4pt}
}
\makeatother

\renewcommand{\paragraph}[1]{\vspace{0.1cm}\noindent{\bf #1.}}

\setcopyright{none} 
\acmConference[Anonymous Submission to ACM CCS 2020]{ACM Conference on Computer and Communications Security}{Due 04 May 2020}{Orlando, FL}
\acmYear{2020}

\begin{document}	
	\hyphenation{Smart-OTPs}
	
	\title{SmartOTPs: An Air-Gapped 2-Factor Authentication for Smart-Contract Wallets (Extended Version)}

	\author{Ivan Homoliak$,^\dagger$$^\ddagger$ Dominik Breitenbacher,$^\ddagger$ Ondrej Hujnak,$^\ddagger$ Pieter Hartel,$^\dagger$ Alexander Binder,$^\dagger$ Pawel Szalachowski$^\dagger$}
	\affiliation{
		\institution{$^\dagger$Singapore University of Technology and Design}
		\institution{$^\ddagger$Brno University of Technology}
	}

	\begin{abstract}
		With the recent rise of cryptocurrencies' popularity, the security and management of crypto-tokens have become critical.
		We have witnessed many attacks on users and providers, which have resulted in significant financial losses. 
		To remedy these issues, several wallet solutions have been proposed. 
		However, these solutions often lack either essential security features, usability, or do not allow users to customize their spending rules.
		
		In this paper, we propose SmartOTPs, a smart-contract wallet framework that gives a flexible, usable, and secure way of managing crypto-tokens in a self-sovereign fashion.
		The proposed framework consists of four components (i.e., an authenticator, a client, a hardware wallet, and a smart contract), and it provides 2-factor authentication (2FA) performed in two stages of interaction with the blockchain.
		To the best of our knowledge, our framework is the first one that utilizes one-time passwords (OTPs) in the setting of the public blockchain.
		In SmartOTPs, the OTPs are aggregated by a Merkle tree and hash chains whereby for each authentication only a short OTP (e.g., 16B-long) is transferred from the authenticator to the client.
		Such a novel setting enables us to make a fully air-gapped authenticator by utilizing small QR codes or a few mnemonic words, while additionally offering resilience against quantum cryptanalysis.
		We have made a proof-of-concept based on the Ethereum platform. 
		Our cost analysis shows that the average cost of a transfer operation is comparable to existing 2FA solutions using smart contracts with multi-signatures. 
	\end{abstract}

\maketitle

\section{Introduction}
\label{sec:intro}

The success of cryptocurrencies has surpassed all expectations resulting in various open and decentralized platforms that allow users to conduct monetary transfers, write smart contracts, and participate in predictive markets.  
Cryptocurrencies introduce their own crypto-tokens, which can be transferred in transactions authenticated by private keys that belong to crypto-token owners.
These private keys are managed by a wallet software that gives users an interface to interact with the cryptocurrency.
There are many cases of stolen keys that were secured by various means~\cite{2016-brainwallets,courtois2016speed,CHHMPRSS18,binance-hack-2019}.
Such cases have brought the attention of the research community to the security issues related to key management in cryptocurrencies~\cite{eskandari2018first,goldfeder2015securing,2015-Bitcoin-SOK}.
According to the previous work~\cite{eskandari2018first,2015-Bitcoin-SOK}, there are a few categories of key management approaches. 

In pass\-word-protected wallets, private keys are encrypted with selected passwords.
Unfortunately, users often choose weak passwords that can be brute-forced if stolen by malware~\cite{2015-CCSM-SecureWorks}; optionally, such malware may use a keylogger for capturing a passphrase~\cite{2015-Bitcoin-SOK,2017-keylogger-bc-malware}.
Another similar option is to use password-derived wallets that generate keys based on the provided password.
However, they also suffer from the possibility of weak passwords~\cite{courtois2016speed}. 
Hardware wallets are a category that promises the provision of better security by introducing devices that enable only the signing of transactions, without revealing the private keys stored on the device.
However, these wallets do not provide protection from an attacker with full access to the device~\cite{kraken-trezor-hack,kraken-keepkey-hack,donjon-ellipal-hack}, and more importantly, wallets that do not have a secure channel for informing the user about the details of a transaction being signed (e.g.,~\cite{ledger-nano}) may be exploited by malware targeting IPC mechanisms~\cite{bui2018man}.

A popular option for storing private keys is to deposit them into  server-side hosted (i.e., custodial) wallets and currency-exchange services~\cite{CoinbaseWallet,binance-exchange,poloniex-exchange,kraken-exchange,luno-wallet,paxful-wallet}.
In contrast to the previous categories, server-side wallets imply trust in a provider, which is a potential risk of this category.
Due to many cases of compromising server-side wallets~\cite{2018-coindesk-bithumb,2014-Mt-Gox,2016-Bitfinex-hack,moore2013beware,binance-hack-2019} or fraudulent currency-exchange operators~\cite{vasek2015there}, client-side hosted wallets have started to proliferate.
In such wallets, the main functionality, including the storage of private keys, has moved to the user side~\cite{mycelium-wallet,CarbonWallet,CitoWiseWallet,coinomi-wallet,InfinitoWallet};
hence, trust in the provider is reduced but the users still depend on the provider's infrastructure.

To increase security of former wallet categories, multi-factor authentication (MFA) is often used, which enables spending crypto-tokens only  when a number of secrets are used together.
However, we emphasize that different security implications stem from the multi-factor authentication made \textit{against a centralized party} (e.g., using Google Authenticator) and \textit{against the blockchain} itself.
In the former, the authentication factor is only as secure as the centralized party, while the latter provides stronger security that depends on the assumption of an honest majority of decentralized consensus nodes  (i.e., miners) and security of cryptographic primitives used. 
Wallets from a split control category~\cite{eskandari2018first} provide MFA against the blockchain.
This can be achieved by threshold cryptography wallets~\cite{goldfeder2015securing,mycelium-entropy}, multi-signature wallets~\cite{Armory-SW-Wallet,Electrum-SW-Wallet,TrustedCoin-cosign,copay-wallet}, and state-aware smart-contract wallets~\cite{TrezorMultisig2of3,parity-wallet,ConsenSys-gnosis}.
The last class of wallets is of our concern, as spending rules and security features can be encoded in a smart contract. 

Although there are several smart-contract wallets using MFA against the blockchain~\cite{TrezorMultisig2of3,ConsenSys-gnosis}, to the best of our knowledge, none of them provide an air-gapped authentication in the form of short OTPs similar to Google Authenticator. 

\paragraph{\textbf{Proposed Approach}}
In this paper, we propose SmartOTPs, a framework for smart-contract  cryptocurrency wallets,
which provides 2FA against data stored on the blockchain.
The first factor is represented by the user's private key and the second factor by OTPs.
To produce OTPs, the authenticator device of SmartOTPs utilizes hash-based cryptographic constructs, namely a pseudo-random function, a Merkle tree, and hash chains.
We propose a novel combination of these elements that minimizes the amount of data transferred from the authenticator, which enables us to implement the authenticator in a fully air-gapped setting, not requiring any USB or another connection. 
SmartOTPs belongs to the category of state-aware smart contract wallets, and it provides protection against the attacker that possesses the user's private key \textit{or} the user's authenticator \textit{or} the attacker that tampers with the client.

\paragraph{\textbf{Contributions}}
Our main contributions are as follows:
\begin{compactitem}

    \item We show that standard 2FA methods against the blockchain do not meet either the security or usability requirements for an air-gapped setting (see \autoref{sec:design-space}).

    \item We propose SmartOTPs, a smart-contract wallet framework that provides 2FA against the blockchain while using short OTPs serving as the second factor (see \autoref{sec:overview}).     
    OTPs are managed in a novel way,  enabling us to make an authenticator device fully air-gapped. 
	
	\item To increase the number of OTPs, we resolve the time-space trade-off at the client by combining hash chains with Merkle trees in a novel way (see \autoref{sec:increasingNoOfOTPs}).
   
	\item We implement and evaluate our approach (including hardware version of the authenticator), and we provide the source code of our solution (see \autoref{sec:implementation}). 		
\end{compactitem}

\medskip
\noindent
Note that this paper is an extended version of our paper published at ACM AFT 2020~\cite{homoliak2020smartotps}. 
In contrast to it, the current paper contains a more detailed related work to cryptocurrency wallets, proposes a classification scheme for such wallets, and fixes a bug in \autoref{alg:new-parent-tree}.

\section{Background and Preliminaries}
\label{sec:pre}

We assume a generic cryptocurrency of which the blocks of records are stored in an ever-growing public distributed ledger called a \textit{blockchain}, which is by design resistant to modifications. 
In a blockchain, blocks are linked using a cryptographic hash function, and each new block has to be agreed upon by participants running a consensus protocol (i.e., \textit{miners}).
Each block may contain orders transferring crypto-tokens, application codes written in a platform-supported language, and the execution orders of such applications. 
These application codes are referred to as \textit{smart contracts} and can encode arbitrary processing logic (e.g., agreements). 
Interactions between clients and the cryptocurrency system are based on messages called \textit{transactions}, which can contain either orders transferring crypto-tokens or calls of smart contract functions.
All transactions sent to a blockchain are validated by miners who replicate the state of the blockchain.

\paragraph{Merkle Tree}\label{sec:MT-background}
A Merkle tree is a data structure based on the binary tree in which every leaf node contains a hash of a single data block, while every non-leaf node contains a hash of its concatenated children.
A Merkle tree enables efficient verification as to whether some data are associated with a leaf node by comparing the expected root hash of a tree with the one computed from a hash of the data in the query and the remaining nodes required to reconstruct the root hash (i.e., \textit{proof} or \textit{authentication path}).
The reconstruction of the root hash has logarithmic time complexity, which makes the Merkle tree an efficient scheme for membership verification.

\subsection{Notation}\label{sec:notation}
By the term \textit{operation} we refer to an action with a smart-contract wallet using SmartOTPs, which may involve, for instance, a transfer of crypto-tokens or a change of daily spending limits.
Then, we use the term \textit{transfer} for the indication of transferring crypto-tokens. 
By $\{msg\}_\mathbb{U}$ we denote the message $msg$ digitally signed by $\mathbb{U}$, and by $msg.\sigma$ we refer to the signature;
$\mathcal{RO}$ is the random oracle;
$h(.)$: stands for a cryptographic hash function;
$h^i(.)$ substitutes $i$-times chained function $h(.)$, e.g., $h^2(.) \equiv h(h(.))$;
$\|$ is the string concatenation;
$h_{\mathcal{D}}^i(.)$ substitutes $i$-times chained function $h(.)$ with embedded domain separation, e.g., $h_{\mathcal{D}}^2(.) = h(2 ~||~ h(1~||~.))$;
$F_k(.) \equiv h(k ~\|~ .)$ denotes a pseudo-random function that is parametrized by a secret seed $k$;
$\%$ represents modulo operation over integers; 
$\Sigma. \{KeyGen, Verify, Sign\}$ represents a signature scheme of the blockchain platform;
$SK_\mathbb{U}$, $PK_\mathbb{U}$ is the private/public key-pair of $\mathbb{U}$, under $\Sigma$,
and $a~|~b$ represents bitwise OR of arguments $a$ and $b$.

\section{Problem Definition}
The main goal of this work is to propose a cryptocurrency wallet framework that provides a secure and usable way of managing crypto-tokens. 
In particular, we aim to achieve:
\begin{compactdesc}
\item[Self-Sovereignty:] 
 ensures that the user does not depend on the 3rd party's infrastructure, and the user does not share his secrets with anybody.
 Self-sovereign (i.e., non-custodial) wallets do not pose a single point of failure in contrast to server-side (i.e., custodial) wallets, which when compromised, resulted in huge financial loses~\cite{2018-coindesk-bithumb,2014-Mt-Gox,2016-Bitfinex-hack,moore2013beware,binance-hack-2019}.
\item[Security:] the insufficient security level of some self-sovereign wallets has caused significant financial losses for individuals and companies~\cite{2016-brainwallets,courtois2016speed,CHHMPRSS18,parity-bug-July-17}. 
We argue that wallets should be designed with security in mind and in particular, we point out 2FA solutions, which have successfully contributed to the security of other environments~\cite{aloul2009two,schneier2005two}.
Our motivation is to provide a cheap security extension of the hardware wallets (i.e., the first factor) by using OTPs as the second factor in a fashion similar to Google Authenticator.   
\end{compactdesc}

\vspace{-0.15cm}
\subsection{Threat Model}\label{sec:threat-model}
For a generic cryptocurrency described in \autoref{sec:pre}, we assume an adversary $\mathcal{A}$ whose goal is to conduct unauthorized operations on the user's behalf or render the user's wallet unusable. 
$\mathcal{A}$ is able to eavesdrop on the network traffic as well as to participate in the underlying consensus protocol. 
However, $\mathcal{A}$ is unable to take over the cryptocurrency platform nor to break the used cryptographic primitives.
We further assume that $\mathcal{A}$ is able to intercept and ``override'' the user's transactions, e.g., by launching a man-in-the-middle (MITM) attack or by creating a conflicting malicious transaction with a higher fee, which will incentivize miners to include $\mathcal{A}$'s transaction and discard the user's one; this attack is also referred to as \textit{transaction front-running}.
We assume three types of exclusively occurring attackers, each targeting one of the three components of our framework: (1) $\mathcal{A}$ with access to the user's private key hardware wallet, (2) $\mathcal{A}$ that tampers with the client, and for completeness we also assume (3) $\mathcal{A}$ with access to the authenticator.
Next, we assume that the legitimate user correctly executes the proposed protocols and $h(.)$ is an instantiation of $\mathcal{RO}$. 

\vspace{-0.15cm}
\subsection{Design Space}\label{sec:design-space}
There are many types of wallets with different properties (see \autoref{sec:soa-wallet-types}).
In our context, to achieve self-sovereignty we identify smart-contract wallets as a promising category.
These wallets manage crypto-tokens by the functionality of smart contracts, enabling users to have customized control over their wallets. 
The advantages of these solutions are that spending rules can be explicitly specified and then enforced by the cryptocurrency platform itself.
Therefore, using this approach, it is possible to build a flexible wallet with features such as daily spending limits or transfer limits.

\paragraph{General OTPs}
With spending rules encoded in a smart contract, it is feasible to design custom security features, such as OTP-based authentication serving as the second factor. 
In such a setting, the authenticator produces OTPs to authenticate transactions in the smart contract.
However, in contrast to digital signatures,
OTPs do not provide non-repudiation of data present in a transaction with an OTP; moreover, they can be intercepted and misused by the front-running or the MITM attacks.
To overcome this limitation, we argue that a two-stage protocol $\Pi_O^{<G>}$ must be employed, enabling secure utilization of general OTPs in the context of blockchains.
In the first stage of $\Pi_O^{<G>}$, an operation $O$, signed by the user $\mathbb{U}$, is submitted to the blockchain, where it obtains an identifier $i$. 
Then, in the second stage, $O_{i}$ is executed on the blockchain upon the submission of $OTP_{i}$ that is unambiguously associated with the operation initiated in the first stage. 

\paragraph{Requirements of General and Air-Gapped OTPs}
Based on the above, we define the necessary security requirements of general OTPs used in the blockchain as follows:
\begin{compactenum}
	\item \textbf{Authenticity:} each OTP must be associated only with a unique authenticator instance. 
	
	\item \textbf{Linkage:} each $OTP_{i}$ must be linked with exactly a single operation $O_i$, ensuring that $OTP_i$ cannot be misused for the authentication of $O_j, i \neq j$.
	
	\item \textbf{Independence:} $OTP_i$ linked with the operation $O_i$
	cannot be derived from $OTP_j$  of an operation $O_j$, where $i \neq j$, or an arbitrary set of other OTPs. 	
\end{compactenum}
Nevertheless, in the air-gapped setting (important for a high usability and security), one more requirement comes into play: \textbf{the short length of OTPs}. 
Short OTPs allow the users to use a relatively small number of mnemonic words or a small QR code to transfer an OTP in an air-gapped fashion.
This requirement is of high importance especially in the case when the authenticator is implemented as a resource-constrained embedded device with a small display (e.g., credit-card-shaped wallet, such as CoolBitX~\cite{CoolWalletS}).

\paragraph{Analysis of Existing Solutions}
We argue that not all solutions meet the requirements of air-gapped OTPs.
Asymmetric cryptography primitives such as digital signatures or zero-knowledge proofs are inadequate in this setting, despite meeting all general OTP requirements.
State-of-the-art signature schemes with reasonable performance overhead~\cite{bernstein2012high,johnson2001elliptic} and short signature size produce a 48B-64B long output.
The BLS signatures~\cite{boneh2001short} go even beyond the previous constructs and might produce signatures of size 32B. 
Nevertheless, BLS signatures are unattractive in the setting of the smart contract platforms that put high execution costs for BLS signature verification, which is $\sim$33 times more expensive than in the case of ECDSA with the equivalent security level~\cite{RFC-BLS-signatures}.
Hence, we assume 48B as the minimal feasible OTP size for assymetric cryptography.

However, transferring even 48B in a fully air-gapped environment by transcription of mnemonic words~\cite{bipMnemonic} would lack usability for regular users -- considering study from Dhakal et al.~\cite{Dhakal-typing-study}, transcription of 36 English words takes 42s on average, which is much longer than users are willing to ``sacrifice.''
We note that the situation is better with QR code, but on the other hand it has two limitations: 
(1) when the authenticator is implemented as a simple embedded device, its display might be unable to fit a requested QR code with sufficient scanning properties (to preserve the maximal scanning distance of QR code, the ``denser'' QR code must be displayed in a larger image~\cite{QR2011size}) and 
(2) occasionally, the users might not have a camera in their devices, thus, they can proceed only with a fallback method that uses mnemonics.
Finally, most of the currently deployed asymmetric constructions are vulnerable to quantum computing~\cite{bernstein2009introduction}.

The problem of long signatures also exists in hash-based signature constructs~\cite{lamport1979constructing-lamportSigs,dods2005hash-winternitz,merkle1989certified}. 
Lamport-Diffie one-time signatures (LD-OTS)~\cite{lamport1979constructing-lamportSigs} produce an output of length $2|h(.)|^2$, which, for example in the case of $|h(.)| = 16B$ yields $4kB$-long signatures. 
The signature size of LD-OTS can be reduced by using one string of one-time key for simultaneous signing of several bits in the message digest (i.e., Winternitz one-time signatures (W-OTS)~\cite{dods2005hash-winternitz}), but at the expense of exponentially increased number of hash computations (in the number of encoded bits) during a signature generation and verification.
The extreme case minimizing the size of W-OTS to $|h(.)|$ (for simplicity omitting checksum) would require $2^{|h(.)|}$ hash computations for signature generation, which is unfeasible. 

Approaches based on symmetric cryptography primitives produce much shorter outputs, but it is challenging to implement them with smart-contract wallets.
Widely used one-time passwords like HOTP~\cite{m2005hotp} or TOTP~\cite{m2011totp} require the user to share a secret key $k$ with the authentication server. 
Then, with each authentication request the user proves that he possesses $k$ by returning the output of an $F_k(.)$ computed with a nonce (i.e., HOTP) or the current timestamp (i.e., TOTP).
This approach is insecure in the setting of the blockchain, as the user would have to share the secret $k$ with a smart-contract wallet, making $k$ publicly visible. 

A solution that does not publicly disclose secret information and, at the same time, provides short enough OTPs (e.g., $16B \simeq 12$ mnemonic words $\simeq$ QR code v1), can be implemented by Lamport's hash chains~\cite{lamport1981password} or other single hash-chain-based constructs, such as T/Key~\cite{kogan2017t}.
A hash chain enables the production of many OTPs by the consecutive execution of a hash function, starting from $k$ that represents a secret key of the authenticator.
Upon the initialization, a smart contract is preloaded with the last generated value $h^n(k)$.
When the user wants to authenticate the $i$th operation, he sends the $h^{n-i}(k)$ to the smart contract in the second stage of $\Pi_O^{<G>}$. 
The smart contract then computes $h(.)$ consecutively $i$ times and checks to ascertain whether the obtained value equals the stored value. 
However, the main drawback of this solution is that \textit{each OTP can be trivially derived from any previous one}, and thereby this scheme does not meet the requirement of OTPs on independence. 
To detail an attack misusing this flaw, assume the MITM attacker possessing $SK_{\mathbb{U}}$ (i.e., the first factor) is able to initiate operations in the first stage of $\Pi_O^{<G>}$.
The attacker $\mathcal{A}$ initiates operation $O_i$ and waits for $\mathbb{U}$ to initiate and confirm an arbitrary follow-up operation $O_j, j > i$. 
When $\mathbb{U}$ sends $OTP_j$ in the second stage of $\Pi_O^{<G>}$, $\mathcal{A}$ intercepts and ``front-runs'' the user's transaction by a malicious transaction with $OTP_i$ computed as $h^{j-i}(OTP_j)$.
Although one may argue that this scheme can be hardened by a modification denying to confirm older operations than the last initiated one, it would bring a race condition issue in which $\mathcal{A}$ might keep initiating operations in the first stage of $\Pi_O^{<G>}$ each time he intercepts a confirmation transaction from $\mathbb{U}$, causing the DoS attack on the wallet.

\section{Proposed Approach}
\label{sec:overview}

For a cryptocurrency described in \autoref{sec:pre}, we propose SmartOTPs, a 2FA against the blockchain, which consists of: (1) a client $\mathbb{C}$, (2) a private key hardware wallet $\mathbb{W}$ equipped with a display, (3) a smart-contract $\mathbb{S}$, and (4) an air-gapped authenticator $\mathbb{A}$ that might be implemented as an embedded device with limited resources.
First, we explain the key idea of our approach, which enables us to construct $\mathbb{A}$ as a fully air-gapped device.
Then, we present the base version of SmartOTPs, and finally, we describe modifications.

\subsection{Design of an Air-Gapped Authenticator}
In our approach, OTPs are generated by a pseudo-random function $F_k(.)$ and then aggregated by a Merkle tree, providing a single value, the root hash ($\mathcal{R}$).
$\mathcal{R}$ is stored at $\mathbb{S}$ and serves as a PK for OTPs.
Assuming the two stage protocol $\Pi_O^{<G>}$ (further denoted as $\Pi_O$), the user $\mathbb{U}$ might confirm the initiated operation $O_{opID}$ by a corresponding $OTP_{opID}$ (provided by $\mathbb{A}$) in the second stage of $\Pi_O$, whereby $\mathbb{S}$ verifies the correctness of $OTP_{opID}$ with use of $\mathcal{R}$.
A challenge of such an approach is the size of an OTP.

\subsubsection{\textbf{From Straw-Man to the Base Version}}
Using the straw-man version, a 2FA requires $\mathbb{A}$ to provide an OTP and its proof.
However, in such a straw-man version, the user $\mathbb{U}$ has to transfer $\frac{(S + S \times H)}{8}$ bytes from $\mathbb{A}$ each time he confirms an operation, where $S$ represents the bit-length of an OTP as well as the output of $h(.)$, and $H$ represents the height of a Merkle tree with $N$ leaves; hence $H = log_2(N$).
For example, if $S = 256$ and $H = 10$, then $\mathbb{U}$ would have to transfer 352B each time he confirms an operation, which has very low usability in an air-gapped setting utilizing transcription of mnemonic words~\cite{bipMnemonic} (i.e., 264 words) or scanning of several QR codes (e.g., 21 QR codes v1) displayed on an embedded device with a small display.
Even further reduction of $S$ to 128 bits would not help to resolve this issue, as the amount of user transferred data would be equal to 176B $\simeq$ 132 mnemonic words $\simeq$ 11 QR codes v1.

We make the observation that it is possible to decouple providing OTPs from providing their proofs.
The only data that need to be kept secret are OTPs, while any node of a Merkle tree may potentially be disclosed -- no OTP can be derived from these nodes.
Therefore, we propose providing OTPs by $\mathbb{A}$, while their proofs can be constructed at $\mathbb{C}$ from stored hashes of OTPs. 
This modification enables us to fetch the nodes of the proof from the storage of $\mathbb{C}$, while $\mathbb{U}$ has to transfer only the OTP itself from $\mathbb{A}$ when confirming an operation (i.e, $S = 128 \simeq 12$ mnemonic words by default).

\subsection{Base Version}\label{sec:base-version}

\begin{algorithm}[t]	
	\caption{Smart contract $\mathbb{S}$ with 2FA}\label{alg:wallet-overview}
	\footnotesize 
	
	\SetKwProg{func}{function}{}{}
	
	$\triangleright$ \textsc{Variables and functions of environment:} \\
	\hspace{1em} \textit{tx}: a current transaction processed by $\mathbb{S}$,\\
	\hspace{1em} \textit{balance}: the current balance of a contract, \\
	\hspace{1em} \textit{transfer(r, v)}: transfer \textit{v} crypto-tokens from a smart contract to \textit{r},\\
	
	\smallskip
	$\triangleright$ \textsc{Declaration of types:}\\
	\hspace{1em} \textbf{Operation} \{ addr, param, pending, type $\in \{ \text{TRANSFER}, \ldots\}$ \} \\
	
	\smallskip
	$\triangleright$ \textsc{Declaration of functions:}
	
	\func{$constructor$(root, pk) \textbf{public} }{
		operations $\leftarrow$ []; \Comment{An append-only list} \\ 
		$PK_\mathbb{U}$ $\leftarrow$ \textit{pk},
		$~$$\mathcal{R}$ $\leftarrow$ root, 
		$~$nextOpID $\leftarrow$ 0; \\
		\textbf{return} $\mathbb{S}^{ID};$ \Comment{Computed by a blockchain platform.} \\
	}

	\func{$initOp$(a, p, type) \textbf{public} }{
		\textbf{assert} $\Sigma.verify(tx.\sigma, PK_\mathbb{U})$; \Comment{1st factor of 2FA} \\						
		opID $\leftarrow$ nextOpID$\texttt{++}$; \\
		operations[opID] $\leftarrow$ \textbf{new} Operation(a, p, \textbf{true}, type); \\
	}
	
	\func{$confirmOp$(otp, $\pi$, opID) \textbf{public} } { 
		\textbf{assert}  operations[opID].pending; \\
		verifyOTP(otp, $\pi$, opID); \Comment{2nd factor of 2FA} \\									
		execOp(operations[opID]); \\
		operations[opID].pending $\leftarrow$ \textbf{false}; \\
	}
	
	\func{$verifyOTP$(otp, $\pi_{opID}$, opID) \textbf{private} } {
		\textbf{assert} deriveRootHash(otp, $\pi_{opID}$, opID) = $\mathcal{R}$; \\
	}
	
	\func{$execOp$(oper) \textbf{private} }{
		\If{TRANSFER = oper.type}{
			\textbf{assert} oper.param  $\leq$ \textit{balance}; \\
			\textit{transfer}(oper.addr, oper.param); \\
			
		} 
	}						
\end{algorithm}

\subsubsection{\textbf{Secure Bootstrapping}}\label{sec:bootstraping-sec}
As common in other schemes and protocols, by default, we assume a secure environment for bootstrapping protocol $\Pi_{B}^\mathcal{S}$ (see \autoref{fig:smart-contract-deploy} and Appendix~\ref{appendix:protocols}). 
First, $\mathbb{A}$ generates a secret seed $k$, which  is stored as a recovery phrase by $\mathbb{U}$.
$\mathbb{W}$ generates a key-pair $SK_\mathbb{U}, PK_\mathbb{U} \leftarrow \Sigma.KeyGen()$.
Next, $\mathbb{U}$ transfers $k$ from $\mathbb{A}$ to $\mathbb{C}$ in an air-gapped manner (i.e., transcribing a few mnemonic words or scanning a QR code). 
Then, $\mathbb{C}$ generates OTPs by computing $F_k(i)~|~i \in \{0, 1, \ldots, N-1\}$, where $N$ is the number of leaves (equal to the number of OTPs in the base version).
Next, $\mathbb{C}$ computes and stores the leaves of the tree -- i.e., the hashes of the OTPs (i.e., $hOTPs$), which do not contain any confidential data.\footnote{To improve performance during provisioning of proofs, $\mathbb{C}$ might additionally store non-leaf nodes, increasing the requirement on $\mathbb{C}$'s storage 2x.}
After this step, $k$ and the OTPs are deleted from $\mathbb{C}$, and $\mathbb{C}$ computes $\mathcal{R}$ from the stored hashes of the OTPs.
Then, $\mathbb{C}$ creates a transaction containing constructor of $\mathbb{S}$ (see \autoref{alg:wallet-overview}) with $\mathcal{R}$ as the argument and passes it to $\mathbb{W}$ for appending $PK_\mathbb{U}$.
Finally, $\mathbb{C}$ sends the transaction with the constructor to the blockchain where the deployment of $\mathbb{S}$ is made.\footnote{$\mathbb{C}$ has the template of $\mathbb{S}$ and the deployment process is unnoticeable for the users.}  
In the constructor, $\mathcal{R}$ with $PK_\mathbb{U}$ are stored and ID of $\mathbb{S}$ (i.e., $\mathbb{S}^{ID}$) is assigned by a blockchain platform and returned in a response.\footnote{Note that $\mathbb{S}^{ID}$ represents a public identification of $\mathbb{S}$, which serves as a destination for sending crypto-tokens to $\mathbb{S}$ by any party.}
Storing $\mathcal{R}$ and $PK_\mathbb{U}$ binds an instance of $\mathbb{S}$ with the user's authenticator $\mathbb{A}$ and the user's private key wallet $\mathbb{W}$, respectively. 
In detail, $PK_\mathbb{U}$ enables $\mathbb{S}$ to verify whether an arbitrary transaction was signed by the user who created $\mathbb{S}$, while $\mathcal{R}$ enables the verification whether the given OTP was produced by the user's $\mathbb{A}$.

\subsubsection{\textbf{Operation Execution}}
\begin{figure}[t]
	\begin{center}
		\vspace{-0.4cm}
		\includegraphics[width=0.47\textwidth]{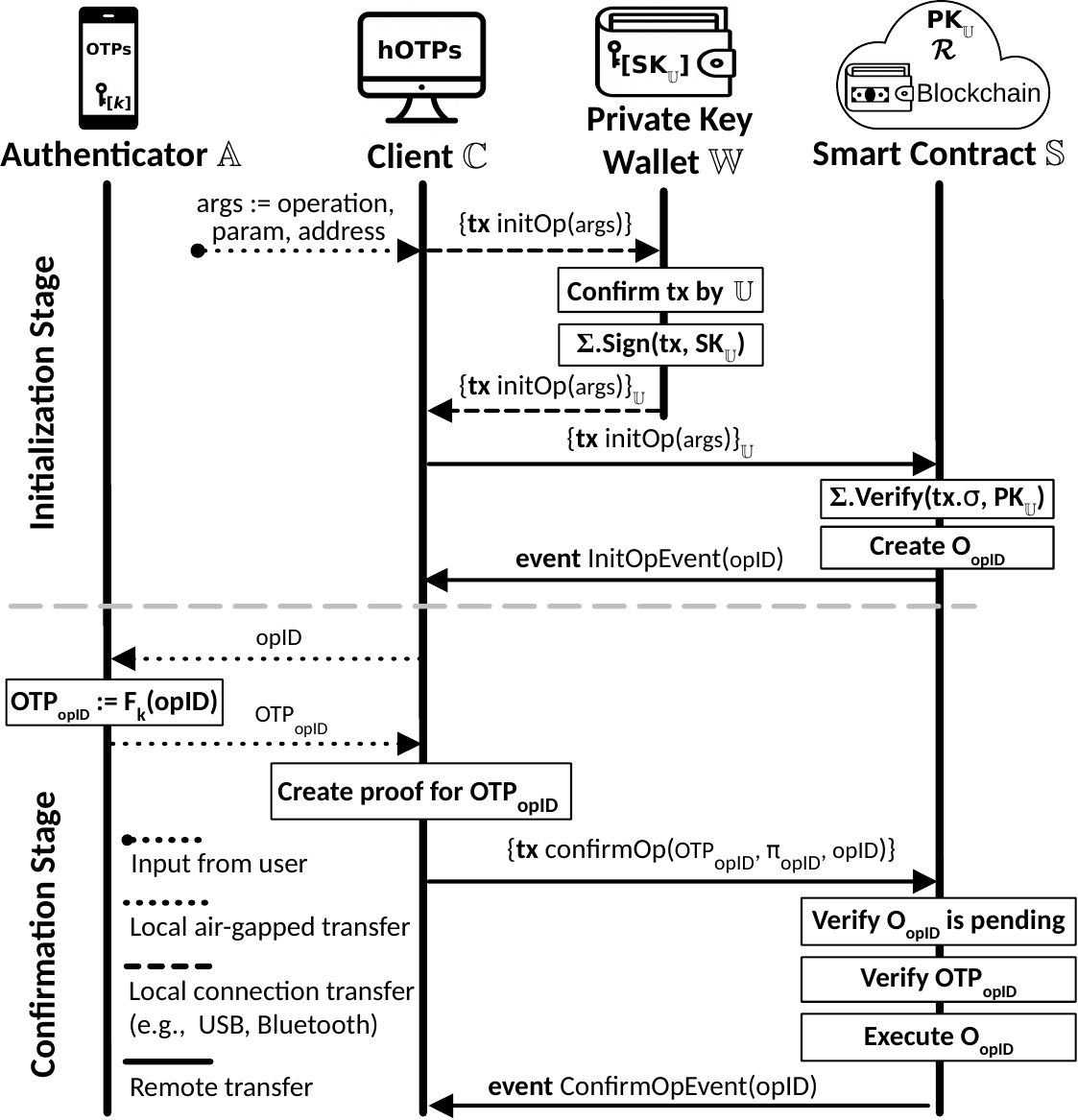} 
		\caption{Execution of an operation ($\Pi_{O}$).}
		\label{fig:smart-contract-execution}		
	\end{center}	
	\vspace{-0.1cm}
\end{figure}

When the wallet framework is initialized, it is ready for executing operations by a two-stage protocol $\Pi_{O}$ (see \autoref{fig:smart-contract-execution} and Appendix~\ref{appendix:protocols}): 
\begin{figure}[t]
	\begin{center}
		\vspace{-0.4cm}
		\includegraphics[width=0.44\textwidth]{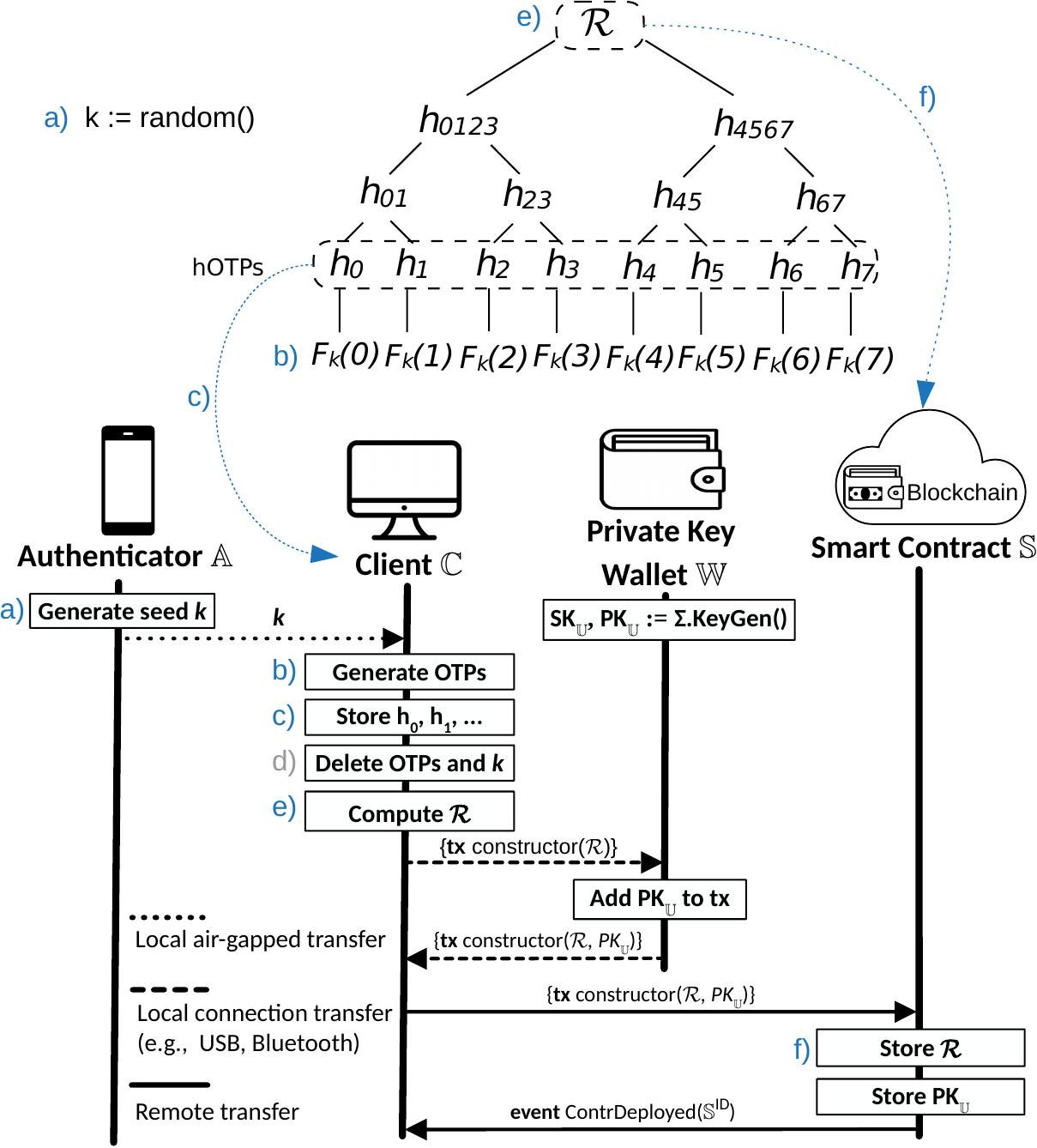} 
		\caption{Bootstrapping of SmartOTPs in a secure environment ($\Pi_{B}^\mathcal{S}$).}
		
		\label{fig:smart-contract-deploy}
		\vspace{-0.1cm}
	\end{center}	
\end{figure}
\begin{compactenum}
    \item \textbf{Initialization Stage}.
	When $\mathbb{U}$ decides to execute an operation with SmartOTPs, he enters the details of the operation into $\mathbb{C}$ that creates a transaction calling \textit{initOp()}, which is provided with operation-specific parameters -- the type of operation (e.g., transfer), a numerical parameter (e.g., amount or daily limit), and an address parameter (e.g., recipient). 
	Then, $\mathbb{C}$ sends this transaction to $\mathbb{W}$, which displays the details of the transaction and prompts $\mathbb{U}$ to confirm signing by a hardware button.
	Upon confirmation, $\mathbb{W}$ signs the transaction by $SK_{\mathbb{U}}$ and sends it back to $\mathbb{C}$.
	$\mathbb{C}$ forwards the transaction to $\mathbb{S}$. 
	In the function \textit{initOp()}, $\mathbb{S}$ verifies whether the signature was created by $\mathbb{U}$ (the first factor), stores the parameters of the operation, and then assigns a sequential ID (i.e., $opID$) to the initiated operation.
	In the response from $\mathbb{S}$, $\mathbb{C}$ is provided with an $opID$.

\item \textbf{Confirmation Stage}.
    After the transaction (that initiated the operation) is persisted on the blockchain, $\mathbb{U}$ proceeds to the second stage of $\Pi_{O}$.
	$\mathbb{U}$ enters $opID$ to $\mathbb{A}$, which, in turn, computes and displays $OTP_{opID}$ as $F_k(opID)$.
    Storing $hOTPs$ computed from OTPs at $\mathbb{C}$ enables $\mathbb{U}$ to transfer only the displayed OTP from $\mathbb{A}$ to $\mathbb{C}$, which can be accomplished in an air-gapped manner. 
    Considering the mnemonic implementation~\cite{bipMnemonic}, this means an air-gapped transfer of 12 words in the case of $O=\text{16B}$.
    Then, $\mathbb{C}$ computes and appends the corresponding proof $\pi_{opID}$ to the OTP.
    The proof of the OTP is computed from stored $hOTPs$ in the $\mathbb{C}$'s storage (or directly fetched from the storage if $\mathbb{C}$ stores all nodes of the Merkle tree).
    Next, $\mathbb{C}$ sends a transaction with $OTP_{opID}$ and its proof $\pi_{opID}$ to the blockchain, calling the function \textit{confirmOp()} of $\mathbb{S}$, which handles the second factor. 
    This function verifies the authenticity of the OTP (i.e., the first requirement of OTPs) and its association with the requested operation (i.e., the second requirement of OTPs), which together implies the correctness of the provided OTP.\footnote{Note that SmartOTPs meet the third requirement of OTPs by the design.}
    In detail, upon calling the \textit{confirmOp()} function with $opID$, $OTP_{opID}$, and $\pi_{opID}$ as the arguments, $\mathbb{S}$ reconstructs the root hash from the provided arguments by the function \textit{deriveRootHash()} that is presented in Appendix~\ref{appendix:algs}.\footnote{Note that this algorithm contains, not yet described, improvements.}
    If the reconstructed value matches the stored value $\mathcal{R}$, the operation is executed (e.g., crypto-tokens are transferred).
    
\end{compactenum}

\noindent
In the following, we present extensions of SmartOTPs, improving its efficiency and usability, and introducing new features.

\subsection{\textbf{Bootstrapping in an Insecure Environment}}\label{sec:bootstraping-insec}
The main advantage of $\Pi_{B}^\mathcal{S}$ described above is its high usability, requiring only an air-gapped transfer of $k$ and connected $\mathbb{W}$.
However, $\Pi_{B}^\mathcal{S}$ is not resistant against $\mathcal{A}$ tampering with $\mathbb{C}$; $\mathcal{A}$ might intercept $k$ or forge $\mathcal{R}$ for $\mathcal{R}'$.
Similarly, $\mathcal{A}$ might forge $PK_\mathbb{U}$ for $PK_\mathcal{A}$, while staying unnoticeable for $\mathbb{U}$ who expects that $\mathbb{S}^{ID}$ obtained is correct.
Therefore, we propose an alternative bootstrapping protocol $\Pi_{B}^\mathcal{I}$ (see Appendix~\ref{appendix:protocols}), assuming that $\mathcal{A}$ can tamper with $\mathbb{C}$ during bootstrapping.
In this protocol, first we protect SmartOTPs from the interception of $k$ and then from forging $\mathcal{R}$ and $PK_\mathbb{U}$.

To avoid the interception of $k$, instead of transferring $k$, $\mathbb{U}$ performs a transfer of all leaves of the Merkle tree (i.e., $hOTPs$) from $\mathbb{A}$ to $\mathbb{C}$, which can be achieved with a microSD card.
Note that the leaves are hashes of OTPs, hence they do not contain any confidential data.
Next, to protect SmartOTPs from forging of $PK_\mathbb{U}$ and $\mathcal{R}$,
we require a deterministic computation of $\mathbb{S}^{ID}$ by a blockchain platform using $PK_\mathbb{U}$ and $\mathcal{R}$, hence $\mathbb{S}^{ID}$ can be computed and displayed together with $\mathcal{R}$ in $\mathbb{W}$ before the deployment of $\mathbb{S}$.
In detail, $\mathbb{S}^{ID}$ is computed as $h(PK_\mathbb{U} ~\|~ \mathcal{R})$, thus each pair consisting of a public key and a root hash  maps to the only $\mathbb{S}^{ID}$.
However, even with this modification, $\mathcal{R}$ can still be forged by $\mathbb{C}$. 
Therefore, when transaction with the constructor is sent to $\mathbb{W}$, $\mathbb{U}$ has to compare $\mathcal{R}$ displayed at $\mathbb{W}$ with the one computed and displayed by $\mathbb{A}$. 
In the case of equality, $\mathbb{U}$ records $\mathbb{S}^{ID}$ displayed in $\mathbb{W}$.

\subsection{Increasing the Number of OTPs}\label{sec:increasingNoOfOTPs}
A small number of OTPs can have negative usability and security implications.
First, users executing many transactions\footnote{E.g., several smart contracts in Ethereum have over $2^{20}$ transactions made.}
would need to create new
OTPs often, and thus change their addresses.
Second, an attacker possessing $SK_{\mathbb{U}}$ can flood $\mathbb{S}$ with initialized operations, rendering all the OTPs unusable.
Therefore, we need to increase the number of OTPs to make the attack unfeasible.
However,  increasing the number of OTPs linearly increases the amount of data
that $\mathbb{C}$ needs to preserve in its storage. 
For example, if the number of OTPs is $2^{20}$, then $\mathbb{C}$ has to
store $33.6MB$ of data (considering $S=16B$ and $\mathbb{C}$ storing all leaves), which is feasible even on storage-limited devices.
However, e.g., for $2^{32}$ OTPs, $\mathbb{C}$ needs to store $137.4GB$ of data, which might be infeasible even on PCs, especially when $\mathbb{C}$ handles multiple instances of SmartOTPs.

To resolve this issue, we modify the base approach by applying a
time-space trade-off~\cite{hellman1980cryptanalytic} for OTPs.
Namely, we introduce  hash chains of which last items are aggregated by the Merkle tree. 
With such a construction, OTPs can be encoded as elements of chains and revealed layer by layer in the reverse order of creating the chains. 
This allows multiplication of the number of OTPs by the chain length without
increasing the $\mathbb{C}$'s storage but imposing a larger number of hash computations on $\mathbb{S}$ and $\mathbb{A}$.
Nonetheless, smart contract platforms set only a low execution cost for $h(.)$.

\begin{figure*}[t]
	\begin{center}
		\vspace{-0.5cm}
		\includegraphics[width=0.98\textwidth]{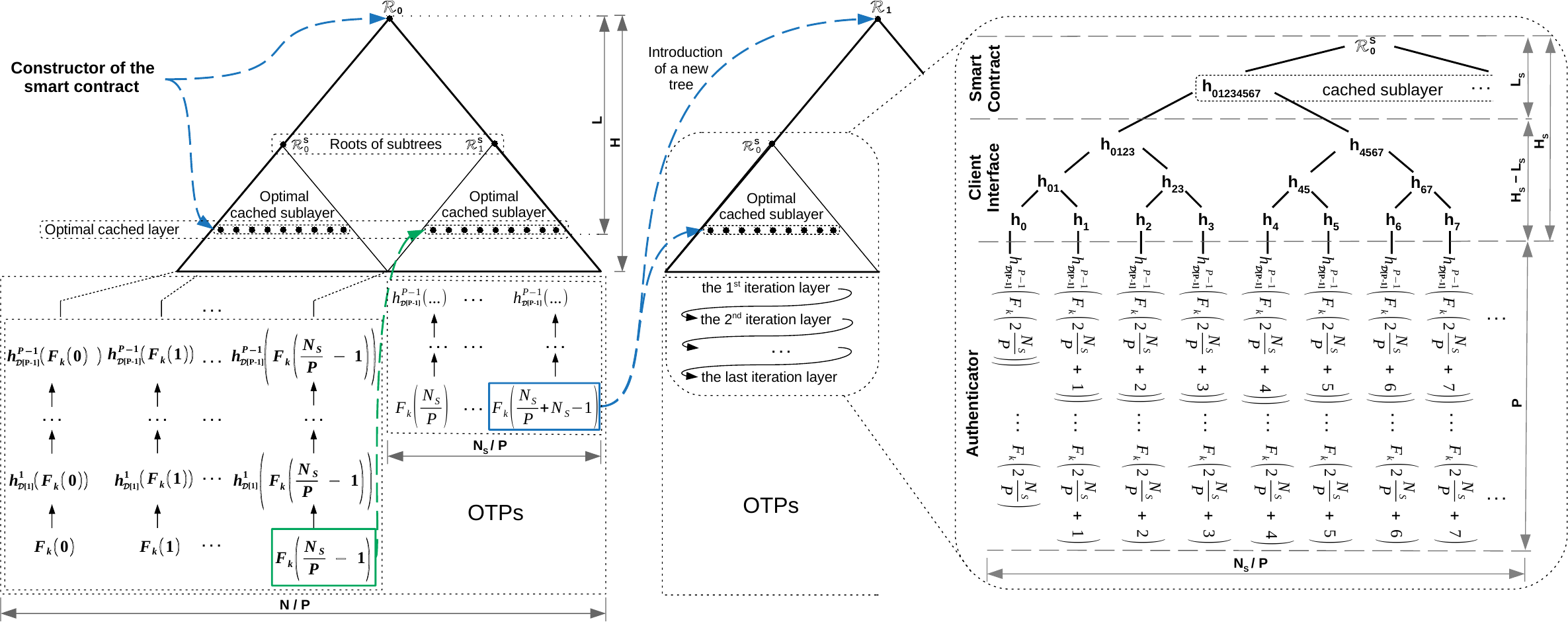} 
		\vspace{-0.4cm}
		\caption{An overview of our approach and its improvements.
		}
		\label{fig:overview}
		\vspace{-0.3cm}
	\end{center}	
\end{figure*}

An illustration of this construction is presented in the bottom left part of
\autoref{fig:overview}.\footnote{Note that this figure contains further, not yet
described, improvements.}
A hash chain of length $P$ is built from each OTP assumed so far. 
Then, the last items of all hash chains are used as the first iteration layer, which provides $\frac{N}{P}$ OTPs.\footnote{For simplicity, we assume  that $GCD(N,P) = P$.} 
Similarly, the penultimate items of all the hash chains are used as the second iteration layer, etc., until the last iteration layer consisting of the first items of hash chains (i.e., outputs of $F_k(.)$) has been reached (see the middle part of \autoref{fig:overview}).
We emphasize that introducing hash chains may cause a violation of the requirement on the independence of OTPs if implemented incorrectly; 
i.e., OTPs from upper iteration layers can be derived from lower layers.
Therefore, to enforce this requirement, we invalidate all the OTPs of all the previous iteration layers by a sliding window at $\mathbb{S}$.

Furthermore, if a hash chain were to use the same hash function throughout the entire chain, it would be vulnerable to birthday attacks~\cite{hu2005efficient}.
To harden a hash chain against a birthday attack, a \textit{domain separation} proposed by Leighton and Micali~\cite{leighton1995large} can be used: a different hash function is applied in each step of a hash chain.
Note that without domain separation, inverting the $i$th iterate of $h(.)$ is $i$ times easier than inverting a single hash function (see the proof in~\cite{haastad2001practical}).
Therefore, we use a different hash function for all but the last iteration layer $1 \leq i < P$ as follows:
\begin{eqnarray}\label{eqn:hash-with-domain-sep}
h_{\mathcal{D}[i]}(x) &=& h(P - i + 1~||~x),
\end{eqnarray}
where $x$ represents the OTP from the next iteration layer.

Although domain separation hardens a single hash chain against the birthday attack, this attack is still possible within the current iteration layer, which is an inevitable consequence of using multiple hash chains.
Therefore, the number of leaves $\mathcal{L}$ (i.e., N/P) is the parameter that must be considered when quantifying the security level of our scheme (see \autoref{sec:analysis}).

With this improvement, $\mathbb{A}$ is updated to provide OTPs by
\begin{eqnarray}\label{eqn:hash-chains}
	getOTP(i) &=& h^{\alpha(i)}_{\mathcal{D}} \Bigg( F_k \Big(\beta(i) \Big) \Bigg),
\end{eqnarray}
where $i$ is the operation ID, $\alpha(i)$ determines the index in a hash chain, 
and $\beta(i)$ determines the index in the last iteration layer of OTPs.
We provide concrete expressions for $\alpha(i)$ and $\beta(i)$ in \autoref{eqn:alpha-beta-final}, which involves all proposed improvements and optimizations.
A derivation of $\mathcal{R}$ from the OTP at $\mathbb{S}$ needs to be updated as well (see \autoref{alg:MT-deriveRH-hash-chains} in Appendix).
In detail, $\mathbb{S}$ executes $P ~-~ \alpha(i) ~-~ 1 = \left\lfloor \frac{i P}{N} \right\rfloor$ hash computations, which is a complementary number to the number of hash computations at $\mathbb{A}$ with regard to $P$.
Also, $\mathbb{C}$ has to be modified, requiring computation of a proof to use the leaf index relative to the current iteration layer of OTPs (i.e., $i ~\%~ \frac{N}{P})$.

With this improvement, given the number of leaves equal to $2^{20}$ and $P=2^{12}$, $\mathbb{C}$ stores only $33.6MB$ of data and it has $2^{32}$ OTPs available.
On the other hand, this modification implies, on average, the execution of additional $P/2$ hash computations  at $\mathbb{S}$, imposing additional costs.
However, our experiments show the benefits of this approach (see \autoref{sec:costs-analysis}).

\begin{algorithm}[!b]
	\footnotesize
	
	\SetKwProg{func}{function}{}{}
	
	$L_1$ $\leftarrow$ []; \Comment{Items have form $<h(\mathcal{R}^{new} ~\|~ OTP)>$}\\
	$L_2$ $\leftarrow$ []; \Comment{Items have form $<$ $\mathcal{R}^{new}$$>$}\\
	
	\smallskip
	\func{$1\_newRootHash$(hRootAndOTP) \textbf{public}}{
		\textbf{assert} $\Sigma.verify(tx.\sigma, PK_\mathbb{U})$;\\
		\textbf{assert} $nextOpID ~\%~ N = N - 1$; \Comment{The last oper. of tree} \\
		$L_1$.append(hRootAndOTP); \\
	}					
	
	\func{$2\_newRootHash$($\mathcal{R}^{new}$) \textbf{public}}{
		\textbf{assert} $\Sigma.verify(tx.\sigma, PK_\mathbb{U})$;\\
		\textbf{assert} nextOpID$~\%~N = N - 1$; \Comment{The last oper. of tree}\\
		$L_2$.append($\mathcal{R}^{new}$);\\
	}
	
	\func{$3\_newRootHash$(otp, $\pi$) \textbf{public}}{
		\textbf{assert} nextOpID$~\%~N = N - 1$; \Comment{The last oper. of tree} \\
		verifyOTP(otp, $\pi$, nextOpID); \\
		\If{$L_1\text{.len} > LEN_{MAX} ~|~ L_2\text{.len} > LEN_{MAX}$}{			
			$L_1$, $L_2$ $\leftarrow$ [], []; \\
			\textbf{return}; \Comment{To avoid $\mathcal{A}$ DoS-ing $\mathbb{S}$ by gas depletion.} \\
		}

		\For{$\{j \leftarrow 0; ~j < L_1\text{.len} ; ~j\texttt{++}\}$}{
			\For{$\{i \leftarrow 0; ~i < L_2\text{.len} ; ~i\texttt{++}\}$}{
				\If{h($L_2$[i] $\|$ otp) = $L_1$[j]}{
					$\mathcal{R}$ $\leftarrow$ $L_2[i]$; \\
					$L_1$, $L_2$ $\leftarrow$ [], []; \\
					nextOpID++; \\
				}
			}
		}	        
		
	}					
	\caption{Introduction of a new $\mathcal{R}$ in $\mathbb{S}$}\label{alg:new-parent-tree}
	\vspace{-0.1cm}
\end{algorithm}

\subsection{Depletion of OTPs}\label{sec:depletion-of-otps}
Even with the previous modification, the number of OTPs remains bounded, therefore they may be depleted.
We propose handling of depleted OTPs by a special operation that replaces the
current tree with a new one.
To introduce a new tree securely, we propose updating $\mathcal{R}$ value while using the last OTP of the current tree for confirmation.  
Nevertheless, for this purpose we cannot use $\Pi_O$ consisting of two stages, as $\mathcal{A}$ possessing $SK_{\mathbb{U}}$ could be ``faster'' than the user and might initialize the last operation and thus block all the user's funds.
If we were to allow repeated initialization of this operation, then we would create a race condition issue.

To avoid this race condition issue, we propose a protocol $\Pi_{NR}$ that replaces $\mathcal{R}$ during three stages of interaction with the blockchain, which requires two append-only lists $L_1$ and $L_2$ (see \autoref{alg:new-parent-tree}):
\begin{compactenum}
	\item $\mathbb{U}$ enters $OTP_{N-1}$ to $\mathbb{C}$. 
	$\mathbb{C}$ sends  $h(OTP_{N-1} ~\|~ \mathcal{R}^{new})$ to $\mathbb{S}$, which appends it to $L_1$.
	
	\item $\mathbb{C}$ sends $\mathcal{R}^{new}$ to $\mathbb{S}$, which appends it to $L_2$.
	
	\item $\mathbb{C}$ passes $OTP_{N-1}$ with $\pi_{N-1}$ to $\mathbb{S}$, where the first matching entries of $L_1$ and $L_2$\footnote{Note that this order must be preserved, otherwise front-running attack is possible. Acknowledgment belongs to Dionysis Zindros who discovered a swapped order of the list iterations in \autoref{alg:new-parent-tree}, presented in the former version of this paper and ACM AFT'20 version~\cite{homoliak2020smartotps}.} are located to perform the introduction of $\mathcal{R}^{new}$.
	Finally, the lists are cleared for future updates.
\end{compactenum}

\noindent
Locating the first entries in the lists relies on the append-only feature of lists, hence no $\mathcal{A}$ can make the first valid pair of entries in the lists.
Similarly as in $\Pi_{B}$, we propose two variants of $\Pi_{NR}$ intended for secure (i.e., $\Pi_{NR}^\mathcal{S}$) and insecure environment (i.e., $\Pi_{NR}^\mathcal{I}$).
In $\Pi_{NR}^\mathcal{I}$ (see Appendix~\ref{appendix:protocols}), $\mathbb{A}$ must compute and display $h(OTP_{N-1} ~\|~ \mathcal{R}^{new})$ and $\mathcal{R}^{new}$ to enable protection against $\mathcal{A}$ that tampers with $\mathbb{C}$. 
Hence, $\mathbb{U}$ can verify the equality of items displayed at $\mathbb{W}$ with the ones displayed at $\mathbb{A}$ during the first and the second stage of $\Pi_{NR}^{\mathcal{I}}$, preventing $\mathcal{A}$ from forging the tree.
To adapt this improvement at $\mathbb{C}$, $\mathbb{C}$ needs to store all nodes of the new tree.
Therefore, $\mathbb{U}$ provides $\mathbb{C}$ with all nodes of the new tree, transferred from $\mathbb{A}$ on a microSD card.
In the case of $\Pi_{NR}^\mathcal{S}$, the nodes of the new tree are transferred by a transcription of $k$ from $\mathbb{A}$ to $\mathbb{C}$ and no values are displayed at $\mathbb{W}$ and $\mathbb{A}$ for $\mathbb{U}$'s verification.

\subsection{Cost \& Security Optimizations }\label{subsec:caching-smart-contract}

\subsubsection{\textbf{Caching in the Smart Contract}}\label{sec:caching-smart-contract}
With a high Merkle tree, the reconstruction of $\mathcal{R}$ from a leaf node may be costly.
Although the number of hash computations stemming from the Merkle tree is logarithmic in the number of leaves, the cost imposed on the blockchain platform may be significant for higher trees.
We propose to reduce this cost by caching an arbitrary tree layer of depth $L$ at $\mathbb{S}$ and do proof verifications against a cached layer.
Hence, every call of \textit{deriveRootHash()} will execute $L$ fewer hash computations in contrast to the version that reconstructs $\mathcal{R}$, while $\mathbb{C}$ will transfer by $L$ fewer elements in the proof. 

The minimal operational cost can be achieved by directly caching leaves of the tree, which accounts only for hash computations coming from hash chains, not a Merkle tree.
However, storing such a high amount of cached data on the blockchain is
too expensive. 
Therefore, this cost optimization must be viewed as a trade-off between the depth $L$ of the cached layer and the price required for the storage of such a cached layer on the blockchain (see
\autoref{sec:costs-analysis}). 

We depict this modification in the left part of \autoref{fig:overview}, and we show that an optimal caching layer can be further partitioned into caching sublayers of subtrees (introduced later).
To enable this optimization, the cached layer of the Merkle tree must be stored in the constructor of $\mathbb{S}$. 
From that moment, the cached layer replaces the functionality of $\mathcal{R}$,  reducing the size of proofs.
During the confirmation stage of $\Pi_{O}$, an OTP and its proof are used for the reconstruction of a particular node in the cached layer, instead of $\mathcal{R}$.
Then the reconstructed value is compared with an expected node of the cached layer.
The index of an expected node is computed as 
\begin{eqnarray}
 idxInCache(i) &=&	\left\lfloor \left(i ~\%~ \frac{N}{P} \right)  ~/~ 2^{H - L} \right\rfloor, 
\end{eqnarray}
where $i$ is the ID of an operation.

\vspace{-0.1cm}
\subsubsection{\textbf{Partitioning to Subtrees}}
The caching of the optimal layer minimizes the operational costs of SmartOTPs, but on the other hand, it requires prepayment for storing the cache on the blockchain.
If the cached layer were to contain a high number of nodes, then the initial deployment cost could be prohibitively high, and moreover, the user might not deplete all the prepaid OTPs.
On top of that, after revealing the first iteration layer of OTPs, the security of our scheme described so far is decreased by $log_2(N/P)$ bits due to the birthday attack (see \autoref{sec:analysis}) on OTPs. 
Hence, bigger trees suffer from higher security loss than smaller trees. 

To overcome the prepayment issue and to mitigate the birthday attack, we propose partitioning an optimal cached layer to smaller groups having the same size, forming sublayers that belong to subtrees (see the left part of \autoref{fig:overview}).
The obtained security loss is $log_2(N_S/P)$, $N_S \ll N$.

Starting with the deployment of $\mathbb{S}$, the cached sublayer of the first subtree and the ``parent'' root hash (i.e., $\mathcal{R}$) are passed to the constructor; the cached sublayer is stored on the blockchain and its consistency against $\mathcal{R}$ is verified.
Then during the operational stage of $\Pi_{O}$, when confirmation of operation is performed, the passed OTP is verified against an expected node in the cached sublayer of the current subtree, saving costs for not doing verification against $\mathcal{R}$ (see \autoref{alg:MT-deriveNodeInCache-hash-chains} in Appendix).

If the last OTP of the current subtree is reached, then no operation other than the introduction of the next subtree can be initialized (see the green dashed arrow in \autoref{fig:overview}).
We propose a protocol $\Pi_{ST}$ for the introduction of the next subtree (see Appendix~\ref{appendix:protocols} for the detailed description).
Namely, $\mathbb{C}$ introduces the next subtree in a single step by calling a function \textit{nextSubtree()} of $\mathbb{S}$ with the arguments containing:
(1) the last OTP of the current subtree $OTP_{(N_S - 1) + \delta N_S}, ~\delta \in \{1, ~\ldots,~ N/N_S - 1\}$, 
(2) its proof $\pi_{otp}$, 
(3) the cached sublayer of the next subtree, and 
(4) the proof $\pi_{sr}$  of the next subtree's root; all items but OTP are computed by $\mathbb{C}$. 
\begin{algorithm}[t]
	\caption{Introduction of the next subtree at $\mathbb{S}$}\label{alg:alg-next-child-tree}	
	
	\footnotesize
	
	\SetKwProg{func}{function}{}{}
	
	currentSubLayer[]; \Comment{Adjusted in the constructor} \\
	\smallskip		
	
	\func{$nextSubtree$(nextSubLayer, otp, $\pi_{otp}$, $\pi_{sr}$) \textbf{public}}{ 
		\textbf{assert} nextOpID $\%~N$ $\neq N - 1$; \Comment{Not the last op. of parent}\\
		\textbf{assert} nextOpID $\%~N_S$ $ = N_S - 1 $; \Comment{The last op. of subtree}\\
		\textbf{assert} currentSubLayer.len = nextSubLayer.len;\\

		\textbf{assert} deriveRootHash(otp,  $\pi_{otp}$, nextOpID) = $\mathcal{R}$;\\				
		
		currentSubLayer $\leftarrow$  nextSubLayer;\\
		
		$\mathcal{R}^s$ $\leftarrow$ reduceMT(currentSubLayer, currentSubLayer.len);\\
		\textbf{assert} subtreeConsistency($\mathcal{R}^s$, $\pi_{sr}$, $\mathcal{R}$);\\
		nextOpID++;  \Comment{Accounts for this introduction of a subtree}\\			
	}			
\end{algorithm}
The pseudo-code of the next subtree  introduction at $\mathbb{S}$ is shown in \autoref{alg:alg-next-child-tree}. 
The current subtree's cached sublayer is replaced by the new one, which is verified by the function $subtreeConsistency()$
against $\mathcal{R}$ with the use of the passed proof 
$\pi_{sr}$ of the new subtree's root hash $\mathcal{R}^s$.
Note that introducing a new subtree invalidates all initialized yet to be confirmed operations of the previous subtree.

At $\mathbb{A}$, this improvement requires accommodating the iteration over layers of hash chains in shorter periods.
Hence, $\mathbb{A}$ provides OTPs by \autoref{eqn:hash-chains} with the following expressions:
\begin{eqnarray}\label{eqn:alpha-beta-final}
\begin{split}
	\alpha(i) &=& P - \left\lfloor   \frac{(i ~\%~ N_S) P}{N_S} \right\rfloor - 1,\\
	\beta(i) &=&  \left\lfloor \frac{i}{N_S} \right\rfloor \frac{N_S}{P} +  \left( i ~\%~ \frac{N_S}{P} \right),
\end{split}
\end{eqnarray}
where $i$ is an operation ID and $N_S$ is the number of OTPs provided by a single subtree.
We remark, that due to this optimization, the update of a new parent root $\mathcal{R}$ as well as the constructor of $\mathbb{S}$ requires, additionally to \autoref{alg:new-parent-tree} and \autoref{alg:wallet-overview}, the introduction of a cached sublayer of the first subtree (omitted here for simplicity).

\section{Security Analysis}
\label{sec:analysis}
We analyze the security of SmartOTPs and its resilience to attacker models under the assumption of random oracle model $\mathcal{RO}$.

\subsection{Security of OTPs}\label{sec:security-of-otps}
OTPs in our scheme are related to two cryptographic constructs: a list of hash chains and the Merkle tree aggregating their last values. 
In this subsection, we assume an adversary $\mathcal{A}$ who is trying to invert OTPs, and we give a concrete expressions for security of our scheme. 
Since we employ the hash domain separation technique~\cite{leighton1995large} for hash chains, each hash execution can be seen as an execution of an independent hash function.
For such a construction, Kogan et al. give the following upper bound (see Theorem 4.6 in~\cite{kogan2017t}) on the advantage of $\mathcal{A}$ breaking a chain:
\begin{eqnarray}\label{eqn:adv-breaks-single-chain}
Pr[\mathit{\mathcal{A}~breaks~a~chain}]\leq\frac{2Q+2P+1}{2^S},
\end{eqnarray}
where $Q$ is the number of queries that $\mathcal{A}$ can make to $h(.)$, $P$ is the chain length, and $S$ is the bit-length of OTPs (and the output of $h(.)$).  
Kogan et al.~\cite{kogan2017t} proved that inverting a hash chain hardened by the domain separation imposes a loss of security equal to the factor of 2.
Therefore, to make a hardened hash chain as secure as $\lambda\varDash{-} bit$ $\mathcal{RO}$, it is enough to set $S = \lambda + 2$.
E.g., to achieve 128-bit security, $S$ should be equal to 130.

\paragraph{SmartOTPs without Subtrees}
This scheme (see \autoref{subsec:caching-smart-contract}) uses a Merkle tree that aggregates $\mathcal{L}=\frac{N}{P}$ hash chains, where
the chains are created independently of each other; they have the same length and the same number of OTPs.
$\mathcal{A}$ can win by inverting any of the chains; hence, the probability that this scheme is secure is
\begin{eqnarray}\label{eqn:scheme-is-secure}
Pr[\mathit{Scheme~is~secure}] = \bigg(1-\frac{2Q+2P+1}{2^S}\bigg)^\mathcal{L}.
\end{eqnarray}
We can apply the alternative form of Bernoulli's inequality
$(1 - x)^\mathcal{L} \geq 1 - x\mathcal{L},$
where $\mathcal{L} \geq 1$ and $0 \leq x \leq 1$ must hold.
In our case, the input conditions hold since the number of hash chains is always greater than one and the probability that $\mathcal{A}$ breaks a single chain from \autoref{eqn:adv-breaks-single-chain} fits the range of $x$ (i.e.,  $0\leq\frac{2Q+2P+1}{2^S}\leq 1$).
Hence, we lower-bound the probability from \autoref{eqn:scheme-is-secure} as follows:
\begin{eqnarray}\label{eqn:scheme-is-secure-nosubtrees}
Pr[\mathit{Scheme~is~secure}] \geq 1 -\frac{\mathcal{L}(2Q+2P+1)}{2^S}.
\end{eqnarray}
\begin{corollary}
To make SmartOTPs without partitioning into subtrees as secure as $\lambda\varDash{-} bit$ $\mathcal{RO}$, it is enough to set $S = \lambda + 2 + log_2(\mathcal{L})$.
\end{corollary}
\noindent For example, to achieve 128-bit security with $\mathcal{L}=64$ and $P \geq 1$, $S$ should be equal to 136, and thus an OTP can be transferred by one QR code v1 or 13 mnemonic words.

\paragraph{Full SmartOTPs}
The full SmartOTPs scheme contains partitioning into subtrees, in which all leaves of the next subtree ``are visible'' only after depleting OTPs of the current subtree (and using OTPs from the 1st iteration layer of the next subtree). 
This improves the security of our scheme under the assumption that $\mathbb{C}$'s storage is not compromised by $\mathcal{A}$, which is true for $\mathcal{A}$ that possesses $PK_{\mathbb{U}}$ or~$\mathbb{A}$.
Therefore, we replace $\mathcal{L}$ in \autoref{eqn:scheme-is-secure-nosubtrees} for $\mathcal{L_S} = \frac{N_S}{P},$ $N_S \ll N$.
\begin{corollary}
	To make the full scheme of SmartOTPs as secure as $\lambda\varDash{-} bit$ $\mathcal{RO}$, it is enough to set $S = \lambda + 2 + log_2(\mathcal{L_S})$.
\end{corollary}
\noindent Therefore, to achieve 128-bit security with $\mathcal{L} = \frac{N}{N_S} \mathcal{L_S}$, $\mathcal{L_S}= 64$, and $P \geq 1$, $S$ should be equal to 136, and thus an OTP can be transferred by a QR code v1 or 13 mnemonic words.
To achieve the same security with $\mathcal{L_S}=1024$, we need to set $S= 140$, and thus an OTP can be transferred in a QR code v2 or 13 mnemonic words.

\subsection{The Attacker Possessing $SK_{\mathbb{U}}$}

\begin{theorem}
$\mathcal{A}$ with access to $SK_{\mathbb{U}}$ is able to initiate operations by $\Pi_{O}$ but is unable to confirm them.
\end{theorem}

\vspace{-0.2cm}
\begin{proof1}
The security of $\Pi_{O}$ is achieved by meeting all requirements on general OTPs (see \autoref{sec:design-space}).
In detail, the requirement on the \textit{independence} of two different OTPs is satisfied by the definition of $F_k(.) \equiv h(k ~\|~ .)$, where $h(.)$ is instantiated by  $\mathcal{RO}$. 
This is applicable when $P=1$.
However, if $P>1$, then items in previous iteration layers of OTPs can be computed from the next ones.
Therefore, to enforce this requirement, we employ an explicit invalidation of OTPs belonging to all previous iteration layers by a sliding window at $\mathbb{S}$ (see \autoref{sec:increasingNoOfOTPs}).
The requirement on the \textit{linkage} of each $OTP_i$ with operation $O_i$ 
is satisfied due to (1) $\mathcal{RO}$ used for instantiation of $h(.)$ and (2) by the definition of the Merkle tree, preserving the order of its aggregated leaves. 
By meeting these requirements, $\mathcal{A}$ is able to initiate an operation $O_j$ in the first stage of $\Pi_{O}$ but is unable to use an $OTP_i$ intercepted in the second stage of $\Pi_{O}$ to confirm $O_j$, where $j \neq i$.
Finally, the requirement on the \textit{authenticity} of OTPs is ensured by a random generation of $k$ and by anchoring $\mathcal{R}$ associated with $k$ at the constructor of $\mathbb{S}$.
\end{proof1}

\begin{theorem}
	Assuming $\delta \in \{0,\ldots, \frac{N}{N_S} - 2\}$, $\mathcal{A}$ with access to $SK_{\mathbb{U}}$ 
	is unable to deplete all OTPs or misuse a stolen OTP that introduces the $(\delta + 1)$th subtree by $\Pi_{ST}$.
\end{theorem}

\vspace{-0.2cm}
\begin{proof1}
When all but one OTPs of the $\delta$th subtree are depleted, the last remaining operation $O_{(N_S - 1) + \delta N_S}, ~\delta \in \{0, ~\ldots,~ \frac{N}{N_S} - 2\}$ is enforced by $\mathbb{S}$ to be the introduction of the next subtree. 
This operation is executed in a single transaction calling the function $nextSubtree()$ of $\mathbb{S}$ (see \autoref{alg:alg-next-child-tree}) requiring the corresponding $OTP_{(N_S - 1) + \delta N_S}$ that is under control of $\mathbb{U}$; hence $\mathcal{A}$ cannot execute the function to proceed with a further depletion of OTPs in the $(\delta + 1)th$ subtree. 
If $\mathcal{A}$ were to intercept $OTP_{(N_S - 1) + \delta N_S}$ during the execution of $\Pi_{ST}$ by $\mathbb{U}$, he could use the intercepted OTP only for the introduction of the next valid subtree since the function $nextSubtree()$ also checks a valid cached sublayer of the $(\delta+1)$th subtree against the parent root hash $\mathcal{R}$.	
\end{proof1}	

\begin{theorem}
Assuming $\delta = \frac{N}{N_S} - 1$,
$\mathcal{A}$ with access to $SK_{\mathbb{U}}$ is neither able to deplete all OTPs nor introduce a new parent tree nor render SmartOTPs unusable. 
\end{theorem}	

\vspace{-0.2cm}
\begin{proof1}
In contrast to the adjustment of the next subtree, the situation here is more difficult to handle, since the new parent tree cannot be verified at $\mathbb{S}$ against any paramount field. 
If we were to use $\Pi_{O}$ while constraining to the last initialized operation $O_{(N-1) + \eta N}, ~\eta \in \{0,1,\ldots\}$ of the parent tree, then $\mathcal{A}$ could render SmartOTPs unusable by submitting an arbitrary $\mathcal{R}$ in $initOp()$, thus blocking all the funds of the user.
If we were to allow repeated initialization of this operation, then we would create a race condition issue.
Therefore, this operation needs to be handled outside of the protocol $\Pi_{O}$, using two unlimited append-only lists $L_1$ and $L_2$ that are manipulated in three stages of interaction with the blockchain (see \autoref{sec:depletion-of-otps}).
In the first stage, $h(\mathcal{R}^{new} ~\|~ OTP_{(N-1) + \eta N})$ is appended to $L_1$, hence $\mathcal{A}$ cannot extract the value of OTP.
In the second stage, $\mathcal{R}^{new}$ is appended to $L_2$,
and finally, in the third stage, the user reveals the OTP for confirmation of the first matching entries in both lists. 
Although $\mathcal{A}$ might use an intercepted OTP from the third stage for appending malicious arguments into $L_1$ and $L_2$, when he proceeds to the third stage and submits the intercepted OTP to $\mathbb{S}	$, the user's entries will match as the first ones. 
\end{proof1}

\subsection{The Attacker that Tampers with the Client}

\begin{theorem}
If $\mathbb{C}$ is tampered with after $\Pi_{B}$, $\mathbb{U}$ can detect such a situation and prevent any malicious operation from being initialized.
\end{theorem}	

\begin{proof1}
If we were to assume that $\mathbb{W}$ is implemented as a software wallet (or hardware wallet without a display), then $\mathcal{A}$ tampering with $\mathbb{C}$ might also tamper with the $\mathbb{W}$'s software running on the same machine. 
This would in turn enable a malicious operation to be initialized and further confirmed by $\mathbb{U}$, since $\mathbb{U}$ would be presented with a legitimate data in $\mathbb{C}$ and $\mathbb{W}$, while the transactions would contain malicious data.
Therefore, we require that $\mathbb{W}$ is implemented as a hardware wallet with a display, which exposes only signing capabilities, while $SK_{\mathbb{U}}$ never leaves the device (e.g., \cite{trezor-hw-wallet,keep-key,BitLox,ellipal-hw-wallet}).
Due to it, $\mathbb{U}$ can verify the details of a transaction being signed in $\mathbb{W}$ and confirm signing only if the details match the information shown in $\mathbb{C}$ (for $\Pi_{O}$) or $\mathbb{A}$ (for $\Pi_{NR}^{\mathcal{I}}$).	
We refer the reader to the work of Arapinis et al.~\cite{ArapinisGKK19} for the security analysis of hardware wallets with displays. 
\end{proof1}

\begin{theorem}
	If $\mathbb{C}$ is tampered with during an execution of  $\Pi_{B}^{\mathcal{I}}$, $\mathcal{A}$ can neither intercept $k$ nor forge $\mathcal{R}$ nor forge $PK_\mathbb{U}$.
\end{theorem}	

\vspace{-0.2cm}
\begin{proof1}
	When the protocol $\Pi_{B}^{\mathcal{I}}$ is used, instead of an air-gapped transfer of $k$ from $\mathbb{A}$ to $\mathbb{C}$, $\mathbb{U}$ transfers leaves of the Merkle tree by microSD card.
	The leaves represent hashes of OTPs in the base version or the hashes of the last items of hash chains in the full version of SmartOTPs.  
	In both versions, the transferred data do not contain any secrets, hence $\mathcal{A}$ cannot take advantage of intercepting them.
	The next option that $\mathcal{A}$ may seek for is to forge $\mathcal{R}$ for $\mathcal{R}'$ and $PK_\mathbb{U}$ for $PK_\mathcal{A}$, which results in different $\mathbb{S}^{ID}$ than in the case of $\mathcal{R}$ and $PK_\mathbb{U}$, since $\mathbb{S}^{ID}$ is computed as $h(PK_\mathbb{U} ~\|~ \mathcal{R})$.
	While $PK_\mathbb{U}$ is stored at $\mathbb{W}$, the authenticity of $\mathcal{R}$ needs to be verified by $\mathbb{U}$ who compares displays of $\mathbb{A}$ and $\mathbb{W}$.
	Only in the case of equality, $\mathbb{U}$ knows that $\mathbb{S}^{ID}$ displayed in $\mathbb{W}$ maps to legitimate $PK_\mathbb{U}$ and $\mathcal{R}$.
\end{proof1}

\subsection{The Attacker Possessing the Authenticator}
It is trivial to see that $\mathcal{A}$ with access to $\mathbb{A}$ is unable to initialize any operation with SmartOTPs since he does not hold $PK_\mathbb{U}$.

\subsection{Further Properties and Implications}

\paragraph{\textbf{Requirement on Block Confirmations}}
Most cryptocurrencies suffer from long time to finality, potentially enabling the accidental forks, which create parallel inconsistent blockchain views.
On the other hand, this issue is not present at blockchain platforms with fast finality, such as Algorand~\cite{gilad2017algorand}, HoneyBadgerBFT~\cite{miller2016honey}, or StrongChain~\cite{strongchain}. 
In blockchains with long time to finality, overly fast confirmation of an operation may be dangerous, as, if an operation were initiated in an ``incorrect'' view, an attacker holding $SK_{\mathbb{U}}$ would hijack the OTP and reuse it for a malicious operation settled in the ``correct'' view. 
To prevent this threat, the recommendation is to wait for several block confirmations to ensure that an accidental fork has not happened. 
For example, in Ethereum, the recommended number of block confirmations to wait is 12 (i.e., $\sim$3 minutes).
Note that such waiting can be done as a background task of $\mathbb{C}$, hence $\mathbb{U}$ does not have to wait: 
(1) considering that $\mathcal{A}$ possesses $SK_{\mathbb{U}}$, $\mathbb{C}$ can detect such a fork during the wait and resubmit the $initOp()$ transaction, 
(2) in the case of $\mathcal{A}$ tampering with $\mathbb{C}$, no operation can be initialized since $\mathbb{U}$ never signs $\mathcal{A}$'s transaction (due to the hardware wallet), and 
(3) $\mathcal{A}$ possessing $\mathbb{A}$ cannot initialize any operation as well. 

\paragraph{Attacks with a Post Quantum Computer}
Although a resilience to quantum computing (QC) is not the focus of this paper, it is of worthy to note that our scheme inherits a resilience to $QC$ from the hash-based cryptography.
The resilience of our scheme to QC is dependent on the output size of $h(.)$. 
A generic QC attack against $h(.)$ is Grover's algorithm~\cite{grover1996fast}, providing a quadratic speedup in searching for the input of the black box function.
As indicated by Amy et al.~\cite{amy2016estimating}, using this algorithm under realistic assumptions, the security of SHA-3 is reduced from 256 to 166 bits.
Applying these results to OTPs with 128-bit security from examples in \autoref{sec:security-of-otps}, we obtain 98-bits post-QC security. 
Further, when assuming the example with $\mathcal{L}=64$ from \autoref{sec:security-of-otps} and~\cite{amy2016estimating}, to achieve 128-bits of post-QC security, we estimate the length of OTPs to 205-bits.

\section{Realization in Practice}
\label{sec:implementation}

We have selected the Ethereum platform and the Solidity language for the implementation of $\mathbb{S}$,
HTML/JS for DAPP of $\mathbb{C}$, Java for smartphone App of $\mathbb{A}$,
 and Trezor T\&One~\cite{trezor-hw-wallet} for $\mathbb{W}$.
We selected $S=128$ bits, which has practical advantages for an air-gapped $\mathbb{A}$, producing OTPs that are 12 mnemonic words long or a QR code v1 (with a capacity of 17B).
Next, we used SHA-3 with truncated output to 128 bits as $h(.)$. 
We selected the size of $k$ equal to 128 bits, fitting 12 mnemonic words $\simeq$ 1 QR code v1.

So far, we have considered only the crypto-token transfer operation.
However, our proposed protocol enables us to extend the set of operations.
For demonstration purposes, we extended the operation set by supporting daily limits and last resort information (see Appendix \ref{sec:functionality-extension}). 
We also tested our contracts by static/dynamic analysis tools Mythril~\cite{mythrill}, Slither~\cite{slither}, and ContractGuard~\cite{ContractGuard-fuzzer}; none of them detected any vulnerabilities. 
In addition, we made a hardware implementation of $\mathbb{A}$ using Node\-MCU~\cite{node-mcu} equipped with ESP8266 (see Appendix~\ref{appendix:NodeMcu-schema}).
The source code of our implementation and videos are available at~\url{https://www.dropbox.com/sh/gmcz8zt12j7omsf/AADR4LHDOhSlwANnI707gkMda?dl=0}.

\subsection{Analysis of the Costs}
\label{sec:costs-analysis}

Executing smart contracts over blockchain, i.e., performing computations and storing data, has its costs. 
In Ethereum Virtual Machine (EVM), these costs are expressed by the level of execution complexity of particular instructions, referred to as \textit{gas}.
One unit of gas has its market price in GWEI.
In this section, we analyze the costs of our approach using the same bit-length $S$ for $h(.)$ as well as for OTPs.
$S$ significantly influences the gas consumption for storing the cached layer on the blockchain. 
We remark that measured costs can also be influenced by EVM internals (e.g., 32B-long words/alignment).

\subsubsection{\textbf{Costs Related to the Merkle Tree}}\label{sec:costs-MK}
\hfill

\begin{figure}[t]
	\begin{center}		
		\includegraphics[width=0.39\textwidth]{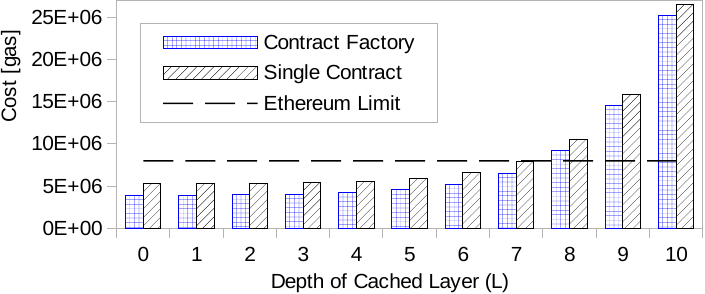} 
		\vspace{-0.3cm}
		\caption{Deployment costs ($H = H_S$).}
		\label{fig:MT-deploy-costs}
		\vspace{-0.1cm}
	\end{center}	
\end{figure}

\paragraph{\textbf{Deployment Cost}}
The cost of a smart contract deployment is driven mainly by the $L_S$ (related to the first subtree) and $S$. 
A less significant factor is the consistency check of a Merkle tree, which is driven by $L_S$: 
the higher $L_S$ is, more layers have to be reduced.
Similarly, the greater $H-H_S$ is, more steps have to be done in the proof verification.
On the other hand, deployment costs are independent of the length $P$ of a hash chain; therefore, we omit the hash chain in this experiment and set $P=1$.
Further, we abstract from the concept of subtrees in order to analyze a single tree (i.e., $H = H_S$).
The deployment costs of our scheme with respect to the depth $L$ ($\equiv L_S$) of the cached layer are presented in \autoref{fig:MT-deploy-costs}.
The figure depicts two cases: one uses a single $\mathbb{S}$ and the second assumes a contract factory producing instances of $\mathbb{S}$.
Thanks to the contract factory, we managed to save a constant amount of  gas equal to $\sim1.3M$, regardless of $L_S$.
Since we assume $8M$ as the maximum gas limit at the Ethereum main network, we can build a caching layer with $L_S=7$ at maximum. 
Later, we will see that the maximum $H_S$ that can be used for the optimal caching layer of a subtree is $H_S = 10$, yielding $2^{10}$ leaves and thus $2^{10} P$ OTPs per subtree.

\begin{figure*}[t]
	\centering	
	\vspace{-0.6cm}
	
	\subfloat[\label{fig:fig:MT-current-benefit-of-caching-256} $H = 8$]{
		\hspace{-0.25cm}
		\includegraphics[width=0.30\textwidth]{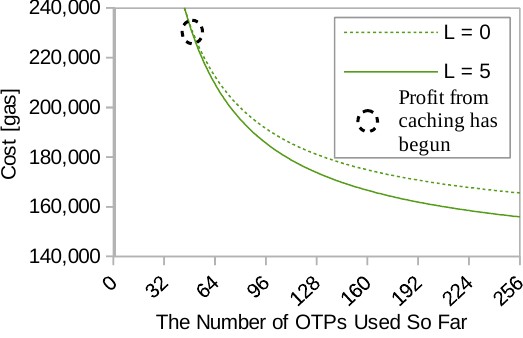} 
	}
	\subfloat[\label{fig:fig:MT-current-benefit-of-caching-512} $H = 9$]{
		\hspace{-0.25cm}
		\includegraphics[width=0.30\textwidth]{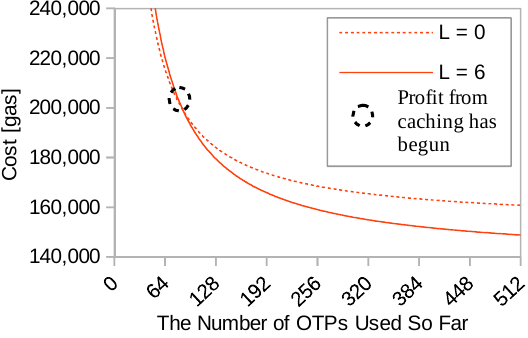} 
	}
	\subfloat[\label{fig:fig:MT-current-benefit-of-caching-1024} $H = 10$]{
		\hspace{-0.25cm}
		\includegraphics[width=0.30\textwidth]{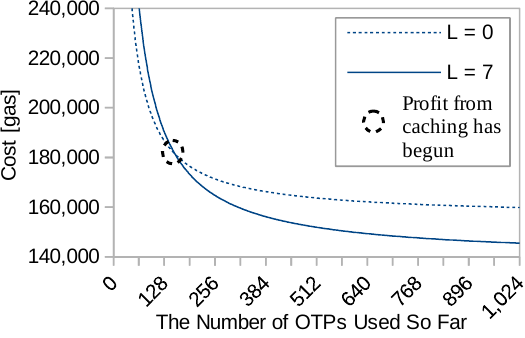} 
	}
	
	\vspace{-0.4cm}
	\caption{Rolling average cost per transfer ($H = H_S$).}
	\label{fig:MT-current-benefit-of-caching}
	\vspace{-0.5cm}	
\end{figure*}

\paragraph{\textbf{Cost of a Transfer}}
Although the cost of each operation supported by $\Pi_{O}$ is similar, here we selected the transfer of crypto-tokens $O^t$, 
and we measured the total cost of $O^t$ as follows: 
\vspace{-0.1cm}
\medmuskip=0mu\thinmuskip=0mu\thickmuskip=-0mu
\begin{eqnarray*}
	O^t\_cost(L, ~N, ~P)&=&  
	  \overline{cost} \left(O^t(L, ~N, ~P) \right)
			    +  \frac{cost \left(O^d(L) \right)}{N}, \\
	\overline{cost} \left( O^t(L, ~N, ~P) \right) &=& \frac{1}{N} \sum_{i=1}^{N}{ cost \left(O^{t}_{i}(L, ~N, ~P) \right)},  \\
	cost \left(O^{t}_{i}(L, ~N, ~P) \right) &=& cost \left( O^{t.init}_i \right) + cost \left( O^{t.confirm}_i(L, ~N, ~P) \right),
\end{eqnarray*}
\medmuskip=3mu\thinmuskip=1mu\thickmuskip=5mu
where \textit{cost()} measures the cost of an operation in gas units, and $O_d$ represents the deployment operation.
As the purpose of the cached layer is to reduce the number of hash computations in \textit{confirmOp()}, the size of an optimal cached layer is subject to a trade-off between the cost of storing the cached layer on the blockchain and the savings benefit of the caching.
To explore the properties of the only Merkle tree, we adjusted $H = H_S$ and $P = 1$.
As each execution of $O^t$ (i.e., $O^t_i$) may have a slightly different gas cost, we measured the average cost of a transaction (i.e., $\overline{cost}(O^t(L, ~N, ~P)$) for both stages of $\Pi_{O}$; note that the cost of $initOp() \simeq 70k$ of gas in all operations.
For completeness, we present the transaction costs of all proposed operations in Appendix~\ref{appendix:cost-of-all-operations}.
In \autoref{fig:MT-expenses}, we can see that the total average cost per transfer decreases with the increasing number of OTPs, as the deployment cost is spread across more OTPs.
The optimal point depicted in the figure minimizes $O^t$ by balancing $cost(O^d(L))$) and  $\overline{cost}(O^t(L, ~N, ~P))$.
We see that $L=H-3$ for such an optimal point.
In contrast to the version without caching, this optimization has brought a cost reduction of $3.87\%, ~5.61\%, ~7.32\%, ~\text{and} ~8.92\%$, for $128$, $256$, $512$, and $1,024$ leaves, respectively.
Next, we explored the number of transfer operations to be executed until a profit of the caching has begun (see \autoref{fig:MT-current-benefit-of-caching}).
We computed a rolling average cost per $O^t$, while distinguishing between the optimal caching layer and disabled caching 
-- the profit from caching begins after $53$, $90$, and $156$ transfers, respectively.

\begin{figure}[t]
	\centering
	\vspace{-0.3cm}
	
	\includegraphics[width=0.35\textwidth]{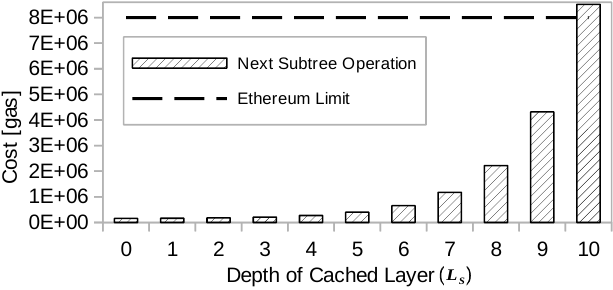} 	
	\vspace{-0.2cm}
	
	\caption{Cost of introducing the next subtree ($H = 20,~ H_S = 10$).}\label{fig:cost-of-new-subtree}
\end{figure}

\paragraph{\textbf{Costs with Subtrees}}
We measured the cost of introducing the next subtree within a parent tree depending on $L_S$, while we set $H = 20$ and $H_S = 10$ (see \autoref{fig:cost-of-new-subtree}).
We found out that when subtrees (and their cached sublayers) are introduced within a dedicated operation, it is significantly cheaper compared to the introduction of a subtree during the deployment.

\begin{figure}[t]
	\centering
	\vspace{-0.3cm}
	\includegraphics[width=0.35\textwidth]{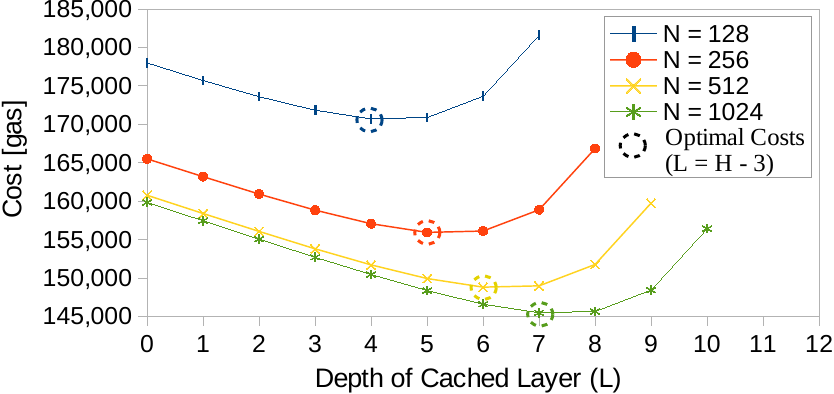} 
	
	\vspace{-0.2cm}
	\caption{Average total cost per transfer ($H = H_S$).}\label{fig:MT-expenses}
\end{figure}

\subsubsection{\textbf{Costs Related to Hash Chains}}
\begin{figure*}[t]
	\centering
	
	\vspace{-0.1cm}
	\hspace{-0.5cm}	
	
	\subfloat[\label{fig:expenses-hash-chain-optim}$L = H$]{
		\hspace{-0.25cm}
		\includegraphics[width=0.30\textwidth]{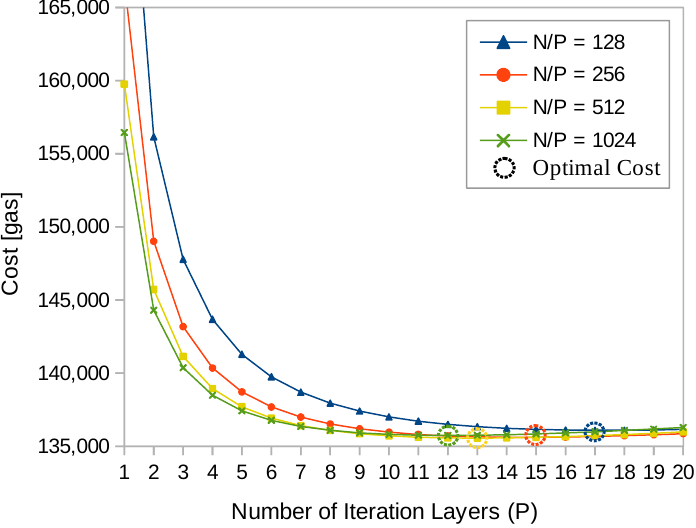} 
	}
	\subfloat[\label{fig:expenses-hash-chain-maxCache}$L = 7$]{
		\hspace{-0.25cm}
		\includegraphics[width=0.285\textwidth]{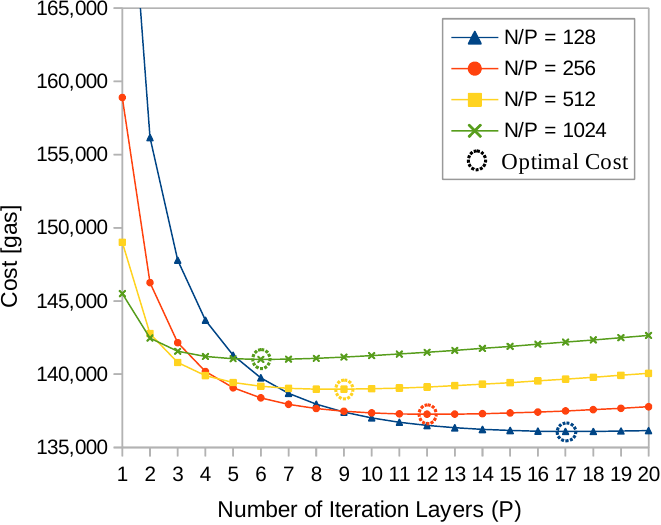} 
	}
	\subfloat[\label{fig:expenses-hash-chain-zero-cache}$L = 0$]{
		\hspace{-0.25cm}
		\includegraphics[width=0.285\textwidth]{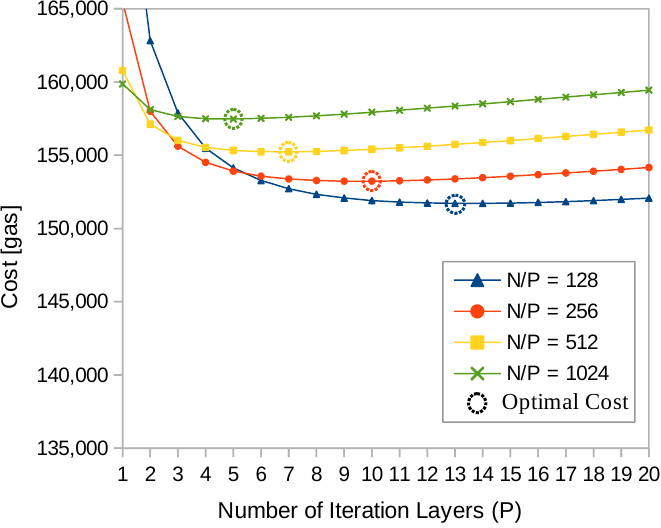} 
	}

	\hspace{1.5cm}
	\vspace{-0.4cm}	
	\caption{Average total cost per transfer with regards to the length $P$ of hash chains.}
	\vspace{-0.3cm}	
	\label{fig:hash-chain-expenses}
\end{figure*}

Since each iteration layer of hash chains contributes to an average cost of $confirmOp()$ with around the same value, we measured this value on a few trees with $P$ up to 512. 
Next, using this value and the deployment cost, we calculated the average total cost per transfer by adding layers of hash chains to a tree with $H = H_S$, thus increasing $N$ by a factor of $P$ until the minimum cost was found.
As a result, the optimal caching layer shifted to the leaves of the tree (see \autoref{fig:expenses-hash-chain-optim}), which would however, exceed the gas limit of Ethereum.
To respect the gas limit, we adjusted $L = 7$, as depicted in \autoref{fig:expenses-hash-chain-maxCache}.
In contrast to the configurations with $L = 0~\text{and}~ P = 1$ (from \autoref{fig:MT-expenses}), we achieved savings of $27.80\%$, $19.61\%$, $14.95\%$, and $12.51\%$ for trees with $H$ equal to $7$, $8$, $9$, and $10$, respectively.
For completeness, we calculated costs for $L=0$ as well (see \autoref{fig:expenses-hash-chain-zero-cache}).
Note that for $L = 0$ and $L = 7$, smaller trees are ``less expensive,'' as they require less operations related to the proof verification in contrast to bigger trees; these operations consume substantially more gas than operations related to hash chains.
Although we minimized the total cost per transfer by finding an optimal $P$, we highlight that increasing $P$ contributes to the cost only minimally but on the other hand, it increases the variance of the cost.
Hence, one may set this parameter even at higher values, depending on the use case.

\subsubsection{\textbf{Costs in Fiat Money}}\label{sec:fiat-money}
We assume the average exchange rate of ETH/USD equal to $211$ and the ``standard'' gas price $5$ GWEI
as of May 2, 2020.
For example, in the case of $N = 2^{25}$ (i.e., $H = 20, ~H_S = 10, ~P = 2^{5}, ~L_S=7 $), expenses per transfer operation are $\$0.2$, while expenses for deployment and introduction of a new subtree are $\$6.90$ and $\$1.23$, respectively.

\vspace{0.1cm}
\section{Related work}
\label{sec:related}
In this section, first we compare SmartOTPs with other hash-based approaches and other smart-contract wallets.
Then, we provide an overview of existing wallet solutions, where we apply and extend the categorization of Eskandari et al.~\cite{eskandari2018first} and Bonneau et al.~\cite{2015-Bitcoin-SOK}.

\paragraph{\textbf{Hash-Based Approaches}}\label{sec:related-comparison}
Although Merkle signatures~\cite{merkle1989certified} utilize Merkle trees for aggregation of several one-time verification keys (e.g.,~\cite{lamport1979constructing-lamportSigs}), the size of these keys and signatures is substantially larger than the size of OTPs in Smart\-OTPs.
Even further optimization of the signature size (i.e., Winternitz OTS~\cite{dods2005hash-winternitz}) does not make signatures as short as in SmartOTPs.
Next, we highlight that we utilize hash chains for multiplication of OTPs, which is different than their application in Winternitz OTS~\cite{dods2005hash-winternitz} that utilize them for the purpose of reducing the size of a single Lamport-Diffie OTS~\cite{lamport1979constructing-lamportSigs} by encoding multiple bits of a message digest into the number of recurrent hash computations.
The next related schemes are Lamport's hash chain~\cite{lamport1981password} and its modification T/Key~\cite{kogan2017t} that applies the domain separation.
However, since they contain only a single chain, they are not secure in the setting of the public blockchain (see \autoref{sec:design-space}) in contrast to SmartOTPs that never consecutively iterate OTPs within a single hash chain.
Moreover, T/Key~\cite{kogan2017t} is using OTPs expiring in 30s to mitigate phishing attacks, which are unrelated in our case.
TESLA~\cite{perrig2000efficient,perrig2005rfc4082} is another related scheme that utilizes a single hash-chain in a centralized setting of time-based multi-cast authentication of streamed messages.


\paragraph{Smart Contract Wallets}
An example of the 2-of-3 multi-signature approach that only supports Trezor wallets is \textit{TrezorMultisig2of3}~\cite{TrezorMultisig2of3}.
A disadvantage of this solution is that $\mathbb{U}$ has to own three Trezor devices, which might be an expensive solution. 
The n-of-m multi-signature scheme is provided by \textit{Gnosis Wallet}~\cite{ConsenSys-gnosis}, which currently holds a significant amount of Ether across various smart contracts.
Similar to the previous example, a disadvantage of this wallet is that $\mathbb{U}$ has to own two hardware wallets for 2FA.

The main reason why existing smart contract wallets using asymmetric cryptography are not suitable for an air-gapped authentication is due to the signature size of 64B.
Hence, to input OTP, $\mathbb{U}$ has to transcribe 48 mnemonic words in the case of lacking a camera on $\mathbb{C}$, which would take $\sim$4x longer than in the case of SmartOTPs.
When $\mathbb{C}$ is equipped with a camera, $\mathbb{A}$ implemented as an embedded device might not be capable of displaying a single OTP as a small QR code since the minimal required QR code having enough data capacity is v4. 
Therefore, several QR codes of a lower version would be needed, which introduces additional complexity for $\mathbb{U}$.

Another drawback of asymmetric cryptography (used in these wallets) stems from its resource demands that increase the operational costs, both on $\mathbb{S}$ and $\mathbb{W}$: 
(1) smart contract platforms place a high execution cost for asymmetric cryptography, and 
(2) $\mathbb{W}$ requires more advanced MCU for cryptographic computations, while $\mathbb{A}$ from SmartOTPs requires only a secure hash function. 
Based on the latter, we believe that hardware realization of $\mathbb{A}$ (see Appendix~\ref{appendix:NodeMcu-schema}) in SmartOTPs is less expensive than the second hardware wallet used in multi-signature smart contracts. 
Moreover, we note that if SmartOTPs were to use only mnemonic words and omit QR codes, then hardware requirements of $\mathbb{A}$ (and thus the overall cost) would be even lower -- mnemonic words can be displayed even on a smart-card-embedded display, such as in CoolBitX~\cite{CoolWalletS}.

\subsection{\textbf{Classification of Authentication Schemes}}\label{sec:classification}
\hfill

\noindent
We introduce the notion of $k$-factor authentication against the blockchain and $k$-factor authentication against the authentication factors.
Using these notions, we propose a classification of authentication schemes, and we apply it to examples of existing key management solutions (see \autoref{sec:soa-wallet-types} and \autoref{appendix:classification}).

In the context of the blockchain, we distinguish between k-factor authentication  \textit{against the blockchain} and k-factor authentication \textit{against the authentication factors} themselves.
For example, an authentication method may require the user to perform 2-of-2 multi-signature in order to execute a transfer, while $\mathbb{U}$ may keep each private key stored in a dedicated device -- each requiring a different password.
In this case, 2FA is performed against the blockchain, since both signatures are verified by all miners of the blockchain.
Additionally, a one-factor authentication is performed once in each device of $\mathbb{U}$ by entering a password in each of them.
For clarity, we classify authentication schemes by the following notation:
\begin{equation*}\label{eqn:auth-simple}
	\Bigg ( 
	Z 
	+  X_1
	\big/ \ldots \big/ 
	X_Z
	\Bigg),
\end{equation*}
where $Z \in \{0, 1, \ldots \}$ represents the number of authentication factors against the blockchain and
$X_i \in \{0, 1, \ldots \} ~|~i \in [1,\ldots, Z]$ represents the number of authentication factors against the i-th factor of $Z$.
With this in mind, we remark that the previous example provides $\left( 2 + 1/1 \right)$-factor authentication: twice against the blockchain (i.e., two signatures), once for accessing the first device (i.e., the first password), and once for accessing the second device (i.e., the second password).
Since the previous notation is insufficient for authentication schemes that use secret sharing~\cite{shamir1979share}, we
extend it as follows:
\begin{equation*}
	\Bigg ( 
	Z^{(W_1, \dots, W_Z)} 
	+ \left( X_1^{1}, \ldots , X_1^{W_1} \right) 
	\big/ \ldots \big/ 
	\left(X_Z^{1}, \ldots, X_Z^{W_Z} \right) 
	\Bigg),
\end{equation*}
where $Z$ has the same meaning as in the previous case, 
$W_i \in \{0, 1, \ldots \}$ $|~i \in [1,\ldots, Z]$ denotes the minimum number of secret shares required to use the complete i-th secret $X_i$.
With this in mind, we remark that the aforementioned example provides $\left( 2^{(1, 1)} + (1)/(1) \right)$-factor authentication: twice against the blockchain (i.e., two signatures), once for accessing the first device (i.e., the first password), and once for accessing the second device (i.e., the second password).
We consider an implicit value of $W_i = 1$; hence, the classification $(2 + 1/1)$ represents the same as the previous one (the first notation suffices).
If one of the private keys were additionally split into two shares, each encrypted by a password, then such an approach would provide $\left( 2^{(2, 1)} + (1, 1)/(1) \right)$-factor authentication.

\subsection{\textbf{Wallet Types}}\label{sec:soa-wallet-types}
\hfill

\noindent
We extend the previous work of Eskandari et al.~\cite{eskandari2018first} and Bonneau et al.~\cite{2015-Bitcoin-SOK}, by categorizing and reviewing a few examples of key management solutions.

\paragraph{\textbf{Keys in Local Storage}}
In this category of wallets, the private keys are stored in plaintext form on the local storage of a machine, thus providing $(1+0)$-factor authentication.
Examples that enable the use of unencrypted private key files are Bitcoin Core~\cite{BitcoinCore} or MyEtherWallet~\cite{MyEtherWallet} wallets.

\paragraph{\textbf{Password-Protected Wallets}}
These wallets require the user-spe\-ci\-fied password to encrypt a private key stored on the local storage, thus providing $(1+1)$-factor authentication.
Examples that support this functionality are Armory Secure Wallet~\cite{Armory-SW-Wallet}, Electrum Wallet~\cite{Electrum-SW-Wallet}, MyEtherWallet~\cite{MyEtherWallet}, Bitcoin Core~\cite{BitcoinCore}, and Bitcoin Wallet~\cite{BitcoinWallet}.
This category addresses physical theft, yet enables the brute force  of passwords and digital theft (e.g., keylogger).

\paragraph{\textbf{Password-Derived Wallets}}
Password-derived wallets~\cite{maxwell2011deterministic} (a.k.a., brain wallets or hierarchical deterministic wallets) can compute a sequence of private keys from only a single mnemonic string and/or password.
This approach takes advantage of the key creation in the ECDSA signature scheme that is used by many blockchain platforms.
Examples of password-derived wallets are Electrum~\cite{Electrum-SW-Wallet}, Armory Secure Wallet~\cite{Armory-SW-Wallet}, Metamask~\cite{MetamaskWallet}, and Daedalus Wallet~\cite{daedalus-wallet}.
The wallets in this category provide $(1+X_1)$-factor authentication (usually $X_1 = 1$) and also suffer from weak passwords~\cite{courtois2016speed}.

\paragraph{\textbf{Hardware Storage Wallets}}
In general, wallets of this category include devices that can only sign transactions by private keys stored inside sealed storage, while the keys never leave the device. 
To sign a transaction, $\mathbb{U}$ connects the device to a machine and enters his passphrase.
When signing a transaction, the device displays the transaction's data to $\mathbb{U}$, who may verify the details. 
Thus, wallets of this category usually provide $(1+1)$-factor authentication.
Popular USB (or Bluetooth) hardware wallets containing displays are offered by Trezor~\cite{trezor-hw-wallet},  Le\-dger~\cite{ledger-nano-s}, KeepKey~\cite{keep-key}, and BitLox~\cite{BitLox}.
An example of a USB wallet that is not resistant against tampering with $\mathbb{C}$ (e.g., keyloggers) is Ledger Nano~\cite{ledger-nano}
-- it does not have a display, hence $\mathbb{U}$ cannot verify the details of transactions being signed.
An air-gapped transfer of transactions using QR codes is provided by ELLIPAL wallet~\cite{ellipal-hw-wallet}.
In ELLIPAL, both $\mathbb{C}$ (e.g., smartphone App) and the hardware wallet must be equipped with cameras and display.
$(1+0)$-factor authentication is provided by a credit-card-shaped hardware wallet from CoolBitX~\cite{CoolWalletS}. 
A hybrid approach that relies on a server providing a relay for 2FA is offered by BitBox~\cite{BitBox}.
Although a BitBox device does not have a display, after connecting to a machine, it communicates with $\mathbb{C}$ running on the machine and at the same time, it communicates with a smartphone App through BitBox's server; 
each requested transaction is displayed and confirmed by $\mathbb{U}$ on the smartphone. 
One limitation of this solution is the lack of self-sovereignty.

\paragraph{\textbf{Split Control -- Threshold Cryptography}}
In threshold cryptography~\cite{shamir1979share,threshold-mackenzie2001two,threshold-gennaro2007secure,blakley1979safeguarding}, a key is split into several parties which enables the spending of crypto-tokens only when n-of-m parties collaborate.
Threshold cryptography wallet provide $\left( 1^{(W_1, \ldots, W_n)}\- + (X_1, \dots, X_n) \right)$-factor authentication, as only a single signature verification is made on a blockchain, but $n$ verifications are made by parties that compute a signature.
Therefore, all the computations for co-signing a transaction are performed off-chain, which provides anonymity of access control policies (i.e., a transaction has a single signature) in contrast to the multi-signature scheme that is publicly visible on the blockchain.
An example of this category is presented by Goldfeder et al.~\cite{goldfeder2015securing}.
One limitation of this solution is a computational overhead that is directly proportional to the number of involved parties $m$ (e.g., for $m = 2$ it takes $13.26$s).
Another example of this category is a USB dongle called Mycelium Entropy~\cite{mycelium-entropy}, which, when connected to a printer, generates triplets of paper wallets using 2-of-3 Shamir's secret sharing; providing $(1^{(2)} + (0, ~0))$-factor authentication.

\paragraph{\textbf{Split Control -- Multi-Signature Wallets}}
In the case of multi-signature wallets, n-of-m owners of the wallet must co-sign the transaction made from the multi-owned address.
Thus, the wallets of this category provide $(n + X_1/\ldots/X_n)$-factor authentication.
One example of a multi-owned address approach is Bitcoin's Pay to Script Hash (P2SH).\footnote{We refer to the term  \textit{multi-owned address of P2SH} for clarity, although it can be viewed as Turing-incomplete smart contract.}
Examples supporting multi-owned addresses are Lockboxes of Armory Secure Wallet~\cite{Armory-SW-Wallet} and Electrum Wallet~\cite{Electrum-SW-Wallet}.
A property of multi-owned address is that each transaction with such an address requires off-chain communication. 
A hybrid instance of this category and client-side hosted wallets category is Trusted Coin's cosigning service~\cite{TrustedCoin-cosign}, which provides a 2-of-3 multi-signature scheme -- $\mathbb{U}$ owns a primary and a backup key, while TrustedCoin owns the third key.
Each transaction is signed first by user's primary key and then, based on the correctness of the OTP from Google Authenticator, by TrustedCoin's key.
Another hybrid instance of this category and client-side hosted wallets is Copay Wallet~\cite{copay-wallet}.
With Copay, the user can create a multi-owned Copay wallet, where $\mathbb{U}$ has all keys in his machines and each transaction is co-signed by n-of-m keys.
Transactions are resent across user's machines during multi-signing through Copay.

\paragraph{\textbf{Split-Control -- State-Aware Smart Contracts}}\label{sec:state-aware=contracts}
State-aware smart contracts provide ``rules'' for how crypto-tokens of a contract can be spent by owners, while they keep the current setting of the rules on the blockchain.
The most common example of state-aware smart contracts is the 2-of-3 multi-signature scheme that provides $(2+X_1/X_2)$-factor authentication.
An example of the 2-of-3 multi-signature approach that only supports Trezor hardware wallets is \textit{TrezorMultisig2of3} from Unchained Capital~\cite{TrezorMultisig2of3}.
One disadvantage of this solution is that $\mathbb{U}$ has to own three Trezor devices, which may be an expensive solution that, moreover, relies only on a single vendor.
Another example of this category, but using the n-of-m multi-signature scheme, is Parity Wallet~\cite{parity-wallet}. 
However, two critical bugs~\cite{parity-bug-July-17,parity-bug-November-17} have caused the multi-signature scheme to be currently disabled.
The n-of-m multi-signature scheme is also used in \textit{Gnosis Wallet} from ConsenSys~\cite{ConsenSys-gnosis}.

\paragraph{\textbf{Hosted  Wallets}}
Common features of hosted wallets are that they provide an online interface for interaction with the blockchain, managing crypto-tokens, and viewing transaction history, while they also store private keys at the server side.
If a hosted wallet has full control over private keys, it is referred to as a \textit{server-side wallet}. 
A server-side wallet acts like a bank -- the trust is centralized.
Due to several cases of compromising such server-side wallets~\cite{2018-coindesk-bithumb}, \cite{2014-Mt-Gox}, \cite{2016-Bitfinex-hack}, \cite{moore2013beware}, the hosted wallets that provide only an interface for interaction with the blockchain (or store only user-encrypted private keys) have started to proliferate.
In such wallets, the functionality, including the storage of private keys, has moved to $\mathbb{U}$'s browser (i.e., client).
We refer to these kinds of wallets as \textit{client-side wallets} (a.k.a., hybrid wallets~\cite{eskandari2018first}).

\paragraph{\textbf{Server-Side Wallets}}
Coinbase~\cite{CoinbaseWallet} is an example of a server-side hosted wallet, which also provides exchange services.
Whenever the user logs in or performs an operation, he authenticates himself against Coinbase's server using a password and obtains a code from Google Authenticator/Authy app/SMS. 
Other examples of server-side wallets having equivalent security level to Coinbase are Circle Pay Wallet~\cite{CircleWallet} and Luno Wallet~\cite{luno-wallet}.
The wallets in this category provide $(0+2)$-factor authentication when 2FA is enabled.

\paragraph{\textbf{Client-Side Wallets}}
An example of a client-side hosted wallet is Blockchain Wallet~\cite{BlockchainInfoWallet}. 
Blockchain Wallet is a password-derived wallet that provides 1-factor authentication against the server based on the knowledge of a password and additionally enables 2FA against the server through one of the options consisting of Google Authenticator, YubiKey, SMS, and email.
When creating a transaction, $\mathbb{U}$ can be authenticated by entering his secondary password. 
Equivalent functionality and security level as in Blockchain Wallet are offered by BTC Wallet~\cite{BTC-com-wallet}. 
In contrast to Blockchain Wallet, BTC wallet uses 2FA also during the confirmation of a transaction. 
Other examples of this category are password-derived wallets, like Mycelium Wallet~\cite{mycelium-wallet}, CarbonWallet~\cite{CarbonWallet}, Citowise Wallet~\cite{CitoWiseWallet}, Coinomi Wallet~\cite{coinomi-wallet}, and Infinito Wallet~\cite{InfinitoWallet}, which, in contrast to the previous examples, do not store backups of encrypted keys at the server.
A 2FA feature is provided additionally to password-based authentication, in the case of CarbonWallet.
In detail, the 2-of-2 multi-signature scheme uses the machine's browser and the smartphone's browser (or the app) to co-sign transactions.

\subsection{\textbf{Classification and Properties of Wallets}}\label{appendix:classification}
\hfill

\noindent
We present a comparison of wallets and approaches from \autoref{sec:soa-wallet-types} in \autoref{tab:state-of-the-art}.
We apply our proposed classification on authentication schemes, while we also survey a few selected security and usability properties of the wallets from the work of Eskandri et al.~\cite{eskandari2018first}. 
In the following, we briefly describe each property and explain the criteria stating how we attributed the properties to particular wallets. 

\paragraph{\textbf{Air-Gapped Property}}
We attribute this property (Y) to approaches that involve at least one hardware device storing secret information, which do not need a connection to a machine in order to operate.

\paragraph{\textbf{Resilience to Tampering with the Client}}
We attribute this property (Y) to all hardware wallets that sign transactions within a device, while they require $\mathbb{U}$ to confirm transaction's details at the device (based on displayed information).
Then, we attribute this property to wallets containing multiple clients that collaborate in several steps to co-signs transactions (a chance that all of them are tampered with is low).

\paragraph{\textbf{Post-Quantum Resilience}}
We attribute this property (Y) to approaches that utilize hash-based cryptography that is known to be resilient against quantum computing attacks~\cite{amy2016estimating}.

\paragraph{\textbf{No Need for Off-Chain Communication}}
We attribute this property (Y) to approaches that do not require an off-chain communication/transfer of transaction among parties/devices
to build a final (co-)signed transaction, before submitting it to a blockchain (applicable only for $Z \geq 2$ or $W_i \ge 2$).

\paragraph{\textbf{Malware Resistance (e.g., Key-Loggers)}}
We attribute this property (Y) to approaches that either enable signing  transactions inside of a sealed device or split signing control over secrets across multiple devices. 

\paragraph{\textbf{Secret(s) Kept Offline}}
We attribute this property (Y) to approaches that keep secrets inside their sealed storage, while they expose only signing functionality.
Next, we attribute this property to paper wallets and fully air-gapped devices.

\paragraph{\textbf{Independence of Trusted Third Party}}
We attribute this property (Y) to approaches that do not require trusted party for operation, while we do not attribute this property to all client-side and server-side hosted wallets.
We partially (P) attribute this property to approaches requiring an external relay server for their operation.

\paragraph{\textbf{Resilience to Physical Theft}}
We attribute this property (Y) to approaches that are protected by an encryption password or PIN.
We partially (P) attribute this property to approaches that do not provide password and PIN protection but have a specific feature to enforce uniqueness of an environment in which they are used (e.g., bluetooth pairing). 

\paragraph{\textbf{Resilience to Password Loss}}
We attribute this property (Y) to approaches that provide means for recovery of secrets (e.g., a seed of hierarchical deterministic wallets).

\setlength{\tabcolsep}{2.0pt}
\begin{table*}[]
	\scriptsize{
		\vspace{-0.3cm}
		\begin{tabular}{lccccccccccccc}
			\toprule
			&
			\multicolumn{2}{c}{\specialcell{\textbf{Authentication Scheme}\\\\}} & \rotatebox[origin=l]{90}{\textbf{\specialcell{Air-Gapped Property}}} &
			\rotatebox[origin=l]{90}{\textbf{\specialcell{Resilience to Tampering\\ with the Client}}} &
			\rotatebox[origin=l]{90}{\textbf{\specialcell{Post-Quantum Resilience}}} &
			\rotatebox[origin=l]{90}{\textbf{\specialcell{No Need for Off-Chain\\Communication}}} &
			\rotatebox[origin=l]{90}{\textbf{\specialcell{Malware Resistance}}} &
			\rotatebox[origin=l]{90}{\textbf{\specialcell{Secret(s) Kept Offline}}} &
			\rotatebox[origin=l]{90}{\textbf{\specialcell{Independence of\\Trusted Third Party}}} &
			\rotatebox[origin=l]{90}{\textbf{\specialcell{Resilience to Physical Theft}}} &
			\rotatebox[origin=l]{90}{\textbf{\specialcell{Resilience to Password Loss}}} &
			\rotatebox[origin=l]{0}{\textbf{\specialcell{Comments}}} \\
			\expandafter\cline\expandafter{\expandafter2\string-3\smallskip}
			&
			\textbf{Classification} &
			\textbf{Details} &
			& & & & & & & & &  \\
			\toprule
			
			\smallskip
			\textbf{Keys in Local Storage}                       & $1 + (0)$          & Private key                                                   &                                               &                                                         &                                                   &                                                   &                                             &                                                  &                                                        &                                                       &                                                      &                                                                                                                  \\
			Bitcoin Core~\cite{BitcoinCore}                                & $1 + (0)$          & For one of the options                                                                      & N                                             & N                                                       & N                                                 & Y                                                 & N                                           & N                                                & Y                                                      & N                                                     & N/A                                                    &                                                                               \\
			MyEtherWallet~\cite{MyEtherWallet}                               & $1 + (0)$          &  For one of the options                                   & N                                             & N                                                       & N                                                 & Y                                                 & N                                           & N                                                & Y                                                      & N                                                     & N/A                                                    & \\
			\midrule
			\smallskip
			\textbf{Password-Protected Wallets}                  & $1 + (1)$          & \specialcell{Private key + encryption}                                  &                                               &                                                         &                                                   &                                                   &                                             &                                                  &                                                        &                                                       &                                                      & \\
			Armory Secure Wallet~\cite{Armory-SW-Wallet}                         & $1 + (1)$          &                                                                       & N                                             & N                                                       & N                                                 & Y                                                 & N                                           & N                                                & Y                                                      & Y                                                     & N                                                    &                                                                               \\
			Electrum Wallet~\cite{Electrum-SW-Wallet}                             & $1 + (1)$          &                                                                       & N                                             & N                                                       & N                                                 & Y                                                 & N                                           & N                                                & Y                                                      & Y                                                     & N                                                    &                                                                               \\
			MyEtherWallet (Offline)~\cite{MyEtherWallet}                     & $1 + (1)$          &                                                                       & N                                             & N                                                       & N                                                 & Y                                                 & N                                           & N                                                & Y                                                      & Y                                                     & N                                                    & \\
			Bitcoin Core~\cite{BitcoinCore}                                 & $1 + (1)$          &                                                                       & N                                             & N                                                       & N                                                 & Y                                                 & N                                           & N                                                & Y                                                      & Y                                                     & N                                                    & \\
			Bitcoin Wallet~\cite{BitcoinWallet}                                & $1 + (1)$          &                                                                       & N                                             & N                                                       & N                                                 & Y                                                 & N                                           & N                                                & Y                                                      & Y                                                     & N                                                    &\\
			\midrule
			\smallskip
			\textbf{Password-Derived Wallets}                    & $1 + (X_1)$         &                                                                       &                                               &                                                         &                                                   &                                                   &                                             &                                                  &                                                        &                                                       &                                                                                           &                                                                               \\
			Armory Secure Wallet~\cite{Armory-SW-Wallet}                         & $1 + (1)$          &                                                                       & N                                             & N                                                       & N                                                 & Y                                                 & N                                           & N                                                & Y                                                      & Y                                                     & Y                                                    & \\
			Electrum Wallet~\cite{Electrum-SW-Wallet}                             & $1 + (1)$          &                                                                       & N                                             & N                                                       & N                                                 & Y                                                 & N                                           & N                                                & Y                                                      & Y                                                     & Y                                                    & \\
			Metamask~\cite{MetamaskWallet}                             & $1 + (1)$          &                                                                       & N                                             & N                                                       & N                                                 & Y                                                 & N                                           & N                                                & Y                                                      & Y                                                     & Y                                                    & \\
			Daedalus Wallet~\cite{daedalus-wallet}                             & $1 + (2)$          & \specialcell{2 passwords}                                          & N                                             & N                                                       & N                                                 & Y                                                 & N                                           & N                                                & Y                                                      & Y                                                     & Y                                                    & \\
			\midrule
			\smallskip
			\textbf{\specialcell{Hardware Storage Wallets}}             &        $1 + (X_1)$        &                                                                       &                                               &                                                         &                                                   &                                                   &                                             &                                                  &                                                        &                                                                                                            &                                      &                                                                               \\
			Trezor~\cite{trezor-hw-wallet}                                      & $1 + (1)$          &                                                                       & N                                             & Y                                                       & N                                                 & Y                                                 & Y                                           & Y                                                & Y                                                      & Y                                                     & Y                                                    & \\
			Ledger~\cite{ledger-nano-s}                                     & $1 + (1)$          &                                                                       & N                                             & Y                                                       & N                                                 & Y                                                 & Y                                           & Y                                                & Y                                                      & Y                                                     & Y                                                    & \\
			KeepKey~\cite{keep-key}                                     & $1 + (1)$          &                                                                       & N                                             & Y                                                       & N                                                 & Y                                                 & Y                                           & Y                                                & Y                                                      & Y                                                     & Y                                                    & \\
			BitLox~\cite{BitLox}                                       & $1 + (2)$          &  2 passwords$^*$                                                                     & N                                             & Y                                                       & N                                                 & Y                                                 & Y                                           & Y                                                & Y                                                      & Y                                                     & Y                                                    & \specialcell{~$^*$Additionally, protection\\~~against the evil maid attack}                                      \\
			CoolWallet S~\cite{CoolWalletS}                       & $1 + (0)$          &                                                                       & N                                             & Y                                                       & N                                                 & Y                                                 & Y                                           & Y                                                & Y                                                      & P$^\dagger$                                                     & N/A                                                    & $^\dagger$\specialcell{Depending on the mode} \\
			
			Ledger Nano~\cite{ledger-nano}                                 & $1 + (2)$          &         Password + GRID card                                                               & N                                             & N                                                       & N                                                 & Y                                                 & N                                           & Y                                                & Y                                                      & Y                                                     & Y                                                    & \\
			ELLIPAL wallet~\cite{ellipal-hw-wallet}                                      & $1 + (1)$          &                                                                       & Y                                             & Y                                                       & N                                                 & Y                                                 & Y                                           & Y                                                & Y                                                      & Y                                                     & Y                                                    & \\
			BitBox USB Wallet~\cite{BitBox}                          & $1 + (2)$          & 1 password and App                                                                       & N                                             & Y                                                       & N                                                 & Y                                                 & Y                                           & Y                                                & P$^\ddagger$                                                      & Y                                                     & Y                                                    & \specialcell{$^\ddagger$Requires a relay server}                \\
			
			\midrule
			\textbf{\specialcell{Split Control --\\Threshold Cryptography}}      & $1^{(W_1)} + (X_1^{1}, \ldots , X_1^{W_1})$         &                                                                       &                                               &                                                         &                                                   &                                                   &                                             &                                                  &                                                        &                                                       &                                                                                            &                                                                               \\
			Goldfeder et al.~\cite{goldfeder2015securing}                            & $1^{(2)} + (1,1)$   & \specialcell{Assuming 2 devices, each\\protected by a password}               & N                                             & Y                                                       & N                                                 & N                                                 & Y                                           & N/A                                              & N/A                                                    & N/A                                                     & N/A &                                                                               \\
			Mycelium Entropy~\cite{mycelium-entropy}                            & $1^{(2)} + (0,0)$   &                                                                       & N                                             & Y                                                       & N                                                 & N                                                 & Y                                           & Y                                                & Y                                                      & Y                                                     & N/A                                                    &                                                                               \\
			\midrule
			\smallskip
			\textbf{\specialcell{Split Control --\\Multi-Signature Wallets}}     & $Z + (X_1/\ldots/X_z)$    &                                                                       &                                               &                                                         &                                                   &                                                   &                                             &                                                  &                                                        &                                                                                                            &                                      &                                                                               \\
			\specialcell{Lockboxes of Armory\\Secure Wallet~\cite{Armory-SW-Wallet}}             & $Z + (X_1/\ldots/X_z)$    & $Z$ up to 7, $X_i$ = 1                                                     & N                                             & Y                                                       & N                                                 & N                                                 & Y                                           & N                                                & Y                                                      & Y                                                     & N                                                    & \\
			Electrum Wallet~\cite{Electrum-SW-Wallet}                             & $Z + (X_1/\ldots/X_z)$    & $Z$ up to 15, $X_i$ = 1                                                    & N                                             & Y                                                       & N                                                 & N                                                 & Y                                           & N                                                & Y                                                      & Y                                                     & Y                                                    & \\
			\specialcell{Trusted Coin's\\cosigning service~\cite{TrustedCoin-cosign}}            &     $2 + (1/2)$           & \specialcell{2 private keys + 2 passwords\\and Google Auth.} & N                                             & Y                                                       & N                                                 & N                                                 & Y                                           & N                                                & N                                                      & Y                                                     & Y                                                    & A hybrid client-side wallet                                                         \\
			Copay Wallet~\cite{copay-wallet}                                & $2 + (1/1)$        &                                                                       & N                                             & Y                                         & N                                                 & N                                                 & Y                                           & N                                                & P                                                      & Y                                                     & Y                                                   & A hybrid client-side wallet                                                          \\
			
			\midrule
			\smallskip
			\textbf{\specialcell{Split-Control --\\State-Aware Smart Contracts}} & $Z + (X_1/\ldots/X_z)$    &                                                                       &                                               &                                                         &                                                   &                                                   &                                             &                                                  &                                                        &                                                       &                                                      & \\
			\specialcell{TrezorMultisig2of3~\cite{TrezorMultisig2of3}}        & $2 + (1/1)$        & \specialcell{Assuming that each device\\is protected by a password}               & N                                             & Y                                                       & N                                                 & N                                                 & Y                                           & Y                                                & Y                                                      & Y                                                     & Y                                                                                &                                                                               \\
			Parity Wallet~\cite{parity-wallet}                               & $Z + (X_1/\ldots/X_z)$    & $Z$ is unlimited, $X_i$ = 1                                                & N                                             & Y                                                       & N                                                 & Y                                                 & Y                                           & N                                                & Y                                                      & Y                                                     & Y                                                    & \\
			Gnosis Wallet~\cite{ConsenSys-gnosis}                              & $Z + (X_1/\ldots/X_z)$    & $Z$ up to 50, $X_i$ = 1                                                    & N                                             & Y                                                       & N                                                 & Y                                                 & Y                                           & N                                                & Y                                                      & N/A                                                   & Y                                                    & \\
			\textbf{\textcolor{blue}{SmartOTPs}}                                        & $2 + (1/1)$        &  \specialcell{Private key and OTPs\\+ passwords}                                                                      & {Y}$^\circ$                                             & Y$^\$$                                                       & {Y}                                                 & Y                                                 & Y                                           & Y                                                & Y                                                      & Y                                                     & 						Y$^{\#}$    								& \specialcell{$^\circ$Fully air-gapped, if combined\\with ELLIPAL wallet\\  $^\$$Thanks to a hardware wallet\\$^{\#}$Also resilient to loss of all secrets} \\
			
			\midrule
			\smallskip
			\textbf{Server-Side Wallets}                         &   $0 + (X_1)$             &                                                                       &                                               &                                                         &                                                   &                                                   &                                             &                                                  &                                                        &                                                       &                                                      &                                      &                                                                               \\
			Coinbase~\cite{CoinbaseWallet}                                    & $0 + (2)$          &    Password, Google Auth./SMS                                                                    & N                                             & N                                                       & N                                                 & Y                                                 & N                                           & N                                                & N                                                      & Y                                                     & Y                                                    & \\
			Circle Pay~\cite{CircleWallet}                           & $0 + (2)$          & \dittoclosing                                                                      & N                                             & N                                                       & N                                                 & Y                                                 & N                                           & N                                                & N                                                      & Y                                                     & Y                                                    & \\
			Luno Wallet~\cite{luno-wallet}                                 & $0 + (2)$          &   Password and Google Auth.                                                                    & N                                             & N                                                       & N                                                 & Y                                                 & N                                           & N                                                & N                                                      & Y                                                     & Y                                                    & \\
			\midrule
			\textbf{Client-Side Wallets}                         &  $Z + (X_1)$          &                                                                       &                                               &                                                         &                                                   &                                                   &                                             &                                                  &                                                        &                                                       &                                                                                          &                                                                               \\
			Blockchain Wallet~\cite{BlockchainInfoWallet}                           & $1 + (2)$          &   \specialcell{Password and one of: Google\\Auth., YubiKey, SMS, or email}                                                                    & N                                             & N                                                       & N                                                 & Y                                                 & N                                           & N                                                & N                                                      & Y                                                     & Y                                                    &  \\
			BTC Wallet~\cite{BTC-com-wallet}                                  & $1 + (2)$          &        \dittoclosing                                 & N                                             & N                                                       & N                                                 & Y                                                 & N                                           & N                                                & N                                                      & Y                                                     & Y                                                    &  \\
			Mycelium Wallet~\cite{mycelium-wallet}                             & $1 + (1)$          &                                                                       & N                                             & N                                                       & N                                                 & Y                                                 & N                                           & N                                                & N                                                      & Y                                                     & Y                                                    &  \\
			CarbonWallet~\cite{CarbonWallet}                                & $2 + (2)$          &   \specialcell{2 private keys stored in\\browser and smartphone}                                                                    & N                                             & Y                                                       & N                                                 & N                                                 & N                                           & N                                                & N                                                      & Y                                                     & Y                                                    & \\
			Citowise Wallet~\cite{CitoWiseWallet}                             & $1 + (2)$          &                                                                       & N                                             & Y$^\P$                                                       & N                                                 & Y                                                 & N                                           & P$^\P$                                                & N                                                      & Y                                                     & Y                                                    &  \specialcell{$^\P$If combined with Trezor\\~or Ledger}               \\
			Coinomi Wallet~\cite{coinomi-wallet}                              & $1 + (1)$          &                                                                       & N                                             & N                                                       & N                                                 & Y                                                 & N                                           & N                                                & N                                                      & Y                                                     & Y                                                    & \\
			Infinito Wallet~\cite{InfinitoWallet}                              & $1 + (1)$          &                                                                       & N                                             & N                                                       & N                                                 & Y                                                 & N                                           & N                                                & N                                                      & Y                                                     & Y                                                    & 
			\\
			\bottomrule
		\end{tabular}
	}
	\caption{Comparison of state-of-the-art cryptocurrency wallets.}	
	\label{tab:state-of-the-art}
\end{table*}
\setlength{\tabcolsep}{1.4pt}

\section{Discussion}
\label{sec:discusion}

\paragraph{\textbf{Vulnerability in HW Wallets}}
We found out that two used hardware wallets do not display all data of transactions being signed: Trezor One displays first 24B of data and Trezor T displays 35B.
With regard to Ethereum transactions, this means that used wallets display only the first 8B and 19B of data representing the parameters of a contract call. 
Hence, $\mathcal{A}$ that tampers with $\mathbb{C}$ might purposely preserve user expected values in the displayed data while forging data that are not displayed. 
We reported this vulnerability to the vendor, and as a mitigation, we put the most critical parameter (i.e., address) of all concerning functions at the first displayed position.

\paragraph{\textbf{Usability}}
Our approach inherits the common usability characteristics of 2FA schemes, such as an extra device to carry,\footnote{Assuming that the user already has a hardware wallet (the first factor).} effort for securely storing the recovery phrase $k$, effort for recalling/entering passwords, and effort for a transfer of OTPs, which can be made by scanning a QR code or transcription of mnemonic words. 
Note that these usability implications are almost the same as in the case of existing smart contract wallets with 2FA~\cite{ConsenSys-gnosis,TrezorMultisig2of3}.
In addition to the previous, SmartOTPs requires $\mathbb{U}$ to introduce a new subtree/parent tree once in a while. 
Nevertheless, we envision this effort to be related only to large businesses rather than regular users; 
considering the example from \autoref{sec:fiat-money}, $\mathbb{U}$ has to introduce the next subtree after using $\sim32K$ OTPs, while $\sim33.5M$ OTPs are available to use before re-initialization of the parent tree.
Next, we note that entering $opID$ into $\mathbb{A}$ might be seen as a usability limitation, especially when $N$ is large.
However, $opID$ can be reset after each iteration layer of the current subtree, thus fitting a small range (i.e., $\langle1,\frac{N_S}{P}\rangle$). 

To compare SmartOTPs with Gnosis Wallet~\cite{ConsenSys-gnosis}, we counted the number of elementary actions (i.e., clicks, button presses, inputs of form fields, QR code scanning) required to make a transfer of funds. 
In the result, SmartOTPs required $33$ actions while Gnosis wallet required $39$ actions.

\paragraph{\textbf{Costs}}
With consumption of up to $\sim150k$ gas units per operation, our approach is comparable to equivalent 2FA solutions using smart contracts: Gnosis Wallet~\cite{ConsenSys-gnosis} requires $\sim275k$ gas units\footnote{\href{https://etherscan.io/tx/0xdb6e9389530a5fa51a43593ac57c3343400fa39ced624d94e266cca4d2b5c09c}{https://etherscan.io/tx/0xdb6e938...} and \href{https://etherscan.io/tx/0x328a7cc75788e0b02a4aa6c7b0ade997c70b3f92522c1ef72f82355bfa2a9455}{https://etherscan.io/tx/0x328a7cc...}} and TrezorMultisig2of3~\cite{TrezorMultisig2of3} requires $\sim95k$ gas units\footnote{\href{https://etherscan.io/tx/0xfc7bbddd8ae58190b5543ed7122e9b52bd67412ffd6b7024e1221d8c80e1229c}{https://etherscan.io/tx/0xfc7bbdd...} (2 signatures in a single transaction).} per operation.

\paragraph{\textbf{Lost Secrets}}
When $\mathbb{U}$ loses access to $\mathbb{A}$, he can initialize a new instance of $\mathbb{A}$ from the backup of seed $k$.
Moreover, if $\mathbb{U}$ losses access to $\mathbb{A}$ and $\mathbb{W}$ at the same time, he can still recover the funds with the last resort functionality that we implemented (see \autoref{sec:implementation}).

\paragraph{\textbf{State at the Client}}
The only state $\mathbb{C}$ has to store is the cache of hashes of OTPs from the first iteration layer. 
This might be seen as a limitation when $\mathbb{U}$ changes a client device. 
However, the state can be recovered anytime from the seed $k$ or transferred by microSD card from $\mathbb{A}$ to $\mathbb{C}$ (see \autoref{sec:bootstraping-sec} and \autoref{sec:bootstraping-insec}).

\paragraph{\textbf{Transaction Size}}
Although the base version of SmartOTPs might slightly bloat the transaction due to $H$ items of the proof, this is improved with the caching at $\mathbb{S}$, which reduces the number of items in the proof to $H$ - $L$.
For example, in the case of $H=10$ and optimal caching (see \autoref{sec:costs-MK}) where $L=7$, only three items of the proof are required.
In this case, SmartOTPs consume ~68B of transaction data (assuming 4B for operation ID), which is similar to asymmetric cryptography used in most of the blockchains.

\section{Conclusion}
\label{sec:conclusion}
In this paper, we have proposed SmartOTPs, a smart-contract wallet framework that provides a secure and usable method of managing crypto-tokens.
The framework provides 2FA that is executed in two stages of interaction with the blockchain and protects against the attacker possessing a user's private key or a user's authenticator or the attacker that tampers with the client.
Our framework uses OTPs constructed using 
a pseudo-random function, Merkle trees, and hash chains.
We combine these primitives in a novel way, which enables an air-gapped setting using transcription of mnemonic words or scanning of small QR codes. 
The protocol of our framework is general and can be utilized, besides the wallets, in any smart contract application for the purpose of 2FA. 
The provided smart contract is self-contained but its operation set can be extended by the community.

\bibliographystyle{ACM-Reference-Format}

\bibliography{ref}

\appendix 
\section{Appendix}\label{sec:appendix}
\medskip

\vspace{-0.2cm}
\subsection{Notation}
We use the following notation in addition to the notation used so far:
$LSB(.)$ extracts a value of the least significant bit; $a \ll b$ represents the bitwise left shift of  $a$ by $b$ bits; 
$a~\&~b$ represents bitwise AND; 
$a~\oplus~b$ represents bitwise exclusive OR; and 
$h_{\mathcal{D}[a:b]}(.)$ represent $(b-a)$-times chained function $h(.)$ with embedded domain separation respecting interval $\langle a, b \rangle$, e.g.,  $h_{\mathcal{D}[2:3]}(.) = h(3 ~\|~ h(2 ~\|~ .)),$

\subsection{Details of Algorithms and Implementation}\label{appendix:algs}
When bootstrapping $\mathbb{C}$, OTPs of the last iteration layer are generated by \autoref{alg:MT-leaves-generation}. 
Generated OTPs are then processed by hash chains, obtaining the first iteration layer of OTPs; this layer is further aggregated into $\mathcal{R}$ by \autoref{alg:MT-reduce-root}, which contains recursive in-situ implementation.
When the $OTP_{opID}$ is used for the authentication of the operation $O_{opID}$, $\mathcal{R}$ is reconstructed from the OTP and its proof $\pi_{opID}$; first, by resolving hash chains and then $\pi_{opID}$ (see \autoref{alg:MT-deriveRH-hash-chains}).

\begin{algorithm}[b]
		\footnotesize
		\caption{Generation of OTPs of the last it. layer}
		\label{alg:MT-leaves-generation}
		
		\SetKwProg{func}{function}{}{}

		\func{$generateOTPs$(k, N, $\eta$)} {
			LL\_OTPs $\leftarrow$ []; \\
			\For{$\{i ~\in~ [0, ~\ldots,~ \frac{N}{P} - 1]\}$} {			
				LL\_OTPs.append($F_{k}(\eta * \frac{N}{P} + ~i)$); 
			}
			\Return LL\_OTPs; \\
		}

\end{algorithm}
\begin{algorithm}[t]
	
	\footnotesize
	\caption{A reconstruction of a node in a cached sublayer of a subtree from OTP and its proof $\pi$}\label{alg:MT-deriveNodeInCache-hash-chains} 
	
	\SetKwProg{func}{function}{}{}
	
	\func{$deriveNodeInCache$(otp, $\pi$, opID) } { 
		\textbf{assert} $\pi$.len = $H_S - L_S$;  \Comment{$H_S = log_2 \left(\frac{N_S}{P}\right)$} \\
		\vspace{-0.1cm}
		
		eci $\leftarrow$ $getExpectedIdxInCache \left(opID ~\%~ \left(\frac{N_S}{P} \right)\right)$; \\
		
		\textbf{assert} eci = deriveIdxInCache($\pi$); \\
		$a \leftarrow \lfloor  (opID ~\%~ N_S)*P / N_S \rfloor$; \\
		res $\leftarrow$ $h_{\mathcal{D}[a:P]}^{a}(otp)$; \Comment{Resolve hash chain} \\
		
		\Comment{Then resolve $\pi$\hfill} \\
		\For{$\{i \leftarrow 0; ~i < \pi.\text{len}; ~\plusplus{i}\}$} { 			
			\If{1 = LSB($\pi$[i])} { 
				res  $\leftarrow$ h(res $\|~ \pi$[i]);   \Comment{A node of $\pi[i]$ is on the right} \\
			} \Else {
				res  $\leftarrow$ h($\pi$[i] $\|$ res);  \Comment{A node of $\pi[i]$ is on the left} \\
			}
		}
		\Return res;
	}
	\vspace{0.05cm}
	
	\func{$deriveIdxInCache$($\pi$)}{ 
		idx $\leftarrow$ 0; \\
		\For{$\{i \leftarrow 0; ~i < H_S - L_S ; ~i\texttt{++}\}$} {
			\If{1 = LSB($\pi$[i])} {
				idx $\leftarrow$ idx $\mathrel{|}$ (1 $\ll$ i); 
			}
		}				         
		\Return idx; \\
	}
	\vspace{0.05cm}
	
	\func{$getExpectedIdxInCache$($childLeafID$)}{ 
		mask $\leftarrow$ 0xFFFFFFFF $\equiv 2^{32} - 1$; \Comment{Assuming max. $H_S = 32$} \\
		retID $\leftarrow$ $childLeafID$; \\
		\For{$\{i \leftarrow H_S - L_S; ~i < H_S ; ~i\texttt{++}\}$} {
			
			bitToClear $\leftarrow$ 0x01 $\ll$ i; \\
			retID $\leftarrow$ retID \& (mask $\oplus$ bitToClear);\\
			
		}				         
		\Return retID; \\
	}					  		
	
\end{algorithm}
\begin{algorithm}[t]
	\footnotesize
	\caption{A reconstruction of $\mathcal{R}$ from OTP and~$\pi$}\label{alg:MT-deriveRH-hash-chains} 
	
	\SetKwProg{func}{function}{}{}
	
	\func{$deriveRootHash$(otp, $\pi$, opID) } { 
		\textbf{assert} $\pi$.len = H;  \Comment{$H = log_2 \left(\frac{N}{P}\right)$} \\
		\vspace{-0.2cm}
		\textbf{assert} $opID ~\%~ \left(\frac{N}{P} \right)$ = deriveIdx($\pi$); \\
		$a \leftarrow \lfloor  (opID ~\%~ N_S)*P / N_S \rfloor$; \\
		res $\leftarrow$ $h_{\mathcal{D}[a:P]}^{a}(otp)$; \Comment{Resolve hash chain} \\
		
		\Comment{Then resolve $\pi$\hfill} \\
		\For{$\{i \leftarrow 0; ~i < \pi.\text{len}; ~\plusplus{i}\}$} { 			
			\If{1 = LSB($\pi$[i])} { 
				res  $\leftarrow$ h(res $\|~ \pi$[i]);   \Comment{A node of $\pi[i]$ is on the right} \\
			} \Else {
				res  $\leftarrow$ h($\pi$[i] $\|$ res);  \Comment{A node of $\pi[i]$ is on the left} \\
			}
		}
		\Return res;
	}
	\vspace{0.05cm}
	
	\func{$deriveIdx$($\pi$)}{ 
		idx $\leftarrow$ 0; \\
		\For{$\{i \leftarrow 0; ~i < \pi.len ; ~i\texttt{++}\}$} {
			\If{1 = LSB($\pi$[i])} {
				idx $\leftarrow$ idx $\mathrel{|}$ (1 $\ll$ i); 
			}
		}				         
		\Return idx; \\
	}			  		
	
\end{algorithm}
\begin{algorithm}[t]
	\footnotesize
	
	\caption{Aggregation of OTPs}\label{alg:MT-reduce-root}
	
	\SetKwProg{func}{function}{}{}
	
	\func{$aggregateOTPs$(OTPs) } {
		hOTPs $\leftarrow$ []; \\
		\For{$\{i ~\in~ [0, ~\ldots,~ OTPs.len - 1]\}$} {
			hOTPs[i] $\leftarrow$ $h^{P}_{\mathcal{D}}(i + 1 ~\|~ OTPs[i])$;	\Comment{Leaves of par. tree} \\
		}
		\Return reduceMT(hOTPs, hOTPs.len); \\
	}
	
	\vspace{0.05cm}	
	\func{$reduceMT$(hashes, length) } {
		\If{1 = length} {
			\Return hashes[0]; \\
		}
		\For{$\{i \leftarrow 0; ~i \leq length / 2; ~i\texttt{++}\}$ } {
			hashes[i] $\leftarrow$ h(hashes[2i] $\|$ hashes[2i + 1]);	\\
		}
		\Return reduceMT(hashes, length / 2); \\
	}
\end{algorithm}

\setlength{\tabcolsep}{4.2pt}
\begin{table}[b]
	\footnotesize
	\centering{
		\begin{tabular}{>{\raggedleft} r r c c c @{}}
			
			\toprule
			\textbf{Operation}                                                                                      & \textbf{Stage} & \textbf{\specialcell{Mean\\$[gas]$}} & \textbf{\specialcell{Standard\\Deviation\\~~~$[gas]$}} & \textbf{\specialcell{Sum\\$[gas]$}}               \\ 
			\Xhline{2\arrayrulewidth} \noalign{\smallskip}
			
			\multirow{2}{*}{\textbf{Transfer}}                                                                      & Init.          & 70,558        & 0                                                                      & \multirow{2}{*}{139,098}   \\
			& Confirm.       & 68,540        & 129                                                                    &                            \\[0.2cm]
			
			\multirow{2}{*}{\textbf{Set Daily Limit}}                                                               & Init.          & 69,342        & 0                                                                      & \multirow{2}{*}{133,938}   \\
			& Confirm.       & 64,596        & 129                                                                    &                            \\[0.2cm]
			
			\multirow{2}{*}{\textbf{\begin{tabular}[c]{@{}r@{}}Set Last  Resort\\ Timeout\end{tabular}}}            & Init.          & 69,342        & 0                                                                      & \multirow{2}{*}{134,324}   \\
			& Confirm.       & 64,982        & 474                                                                    &                            \\[0.2cm]
			
			\multirow{2}{*}{\textbf{\begin{tabular}[c]{@{}r@{}}Set Last  Resort\\ Address\end{tabular}}}            & Init.          & 70,366        & 0                                                                      & \multirow{2}{*}{135,604}   \\
			& Confirm.       & 65,238        & 129                                                                    &                 	\\[0.2cm] 
			
			\multirow{3}{*}{\textbf{\begin{tabular}[c]{@{}r@{}}Introduction \\of the Next \\Parent Tree\end{tabular}}} & Stage 1        & 34,223        & -                                                                      & \multirow{3}{*}{1,165,691} \\
			& Stage 2        & 49,459        & -                                                                      &                            \\
			& Stage 3        & 1,082,009     & -                                                                      & \\[0.2cm] 
			
			\textbf{\begin{tabular}[c]{@{}r@{}}Introduction of\\the Next Subtree\end{tabular}}            & \multicolumn{4}{c}{Depends mainly on $L_S$ (see \autoref{fig:cost-of-new-subtree})}   \\[0.3cm] 
			
			\textbf{\begin{tabular}[c]{@{}r@{}}Send Crypto-Tokens to\\ the Last Resort Address\end{tabular}}            & -              & 13,887        & -                                                                      & 13,887                     \\ 
			\bottomrule
		\end{tabular}
	}
	
	\caption{Costs of all operations ($H = 10,~ L_S = 7, ~P = 1$).}	
	\label{tab:other-costs}
	
\end{table}
\renewcommand{\arraystretch}{1.0}

\subsection{Functionality Extension of the Wallet}\label{sec:functionality-extension}
\paragraph{\textbf{Daily Limit}}
Adjusting a daily  limit is a functionality that contributes primarily to $\mathbb{U}$'s self-monitoring of expenses but at the same time it avoids typos in transfers that exceeds a daily limit. 
This operation has the only argument representing an amount that can be spent in a single calendar day.
Security implications for this operation are the same as in the case of the transfer crypto-tokens operation (see \autoref{sec:analysis}).
\vspace{0.5cm}

\paragraph{\textbf{Last Resort Address and Timeout}}
As users may lose all secrets, leading to an unrecoverable state, we propose an extension that deals with such a situation based on the last resort address and timeout options.
This sort of a functionality needs two dedicated operations of $\Pi_{O}$: one for the adjustment of the last resort address (enforced to be different than the address of $\mathbb{U}$) and another one for the adjustment of the timeout. 
If the timeout has elapsed, then anyone may call a dedicated function that transfers all the funds to the last resort address and destroys the contract.
Note that the last resort address is enforced to be different than the address of the owner of the smart contract in order to avoid transferring all funds of the wallet to the owner's address (i.e., that might be under control of the $\mathcal{A}$) when $\mathbb{U}$ loses all secrets.
Note that update of the activity is made only in the second stage of $\Pi_O$, requiring an OTP.

\subsection{Cost of All Operations}\label{appendix:cost-of-all-operations}
Operational costs of all implemented operations are shown in \autoref{tab:other-costs}.
In the table, we do not account for deployment costs, hence we measure only instant gas consumption of the function calls.
The cost measurements were obtained using configuration with the optimal cost (i.e., $L_S = H_S - 3$), $H = H_S$ and $P = 1$, which are independent of $H$.

\subsection{Detailed Description of Protocols}\label{appendix:protocols}
\vspace{-0.2cm}

\begin{oframed}
	\begin{center}
		\vspace{-0.2cm}
		Bootstrapping -- protocol $\Pi_{B}^\mathcal{S}$
		\\(for a secure environment)
	\end{center}
	
	\begin{asparaitem}
		\item{\textbf{Authenticator $\mathbb{A}$:}
			Generate $k \leftarrow random()$ and display $k$ to $\mathbb{U}$.  
		}
		\item{\textbf{Client $\mathbb{C}$:}
			Upon $k$, $N$, $N_S$, and $P$ are entered by $\mathbb{U}$ into $\mathbb{C}$,  compute $OTPs_{LL} \leftarrow F_k(\eta * \frac{N}{P} + i), ~i \in \{0, \ldots, \frac{N}{P} -1\}, ~\eta \in \{0,1,\ldots \}$. 
			Then compute and store $hOTPs \leftarrow h^{P}_{\mathcal{D}}(OTPs_{LL}[i]), ~i \in \{0, \ldots,  \frac{N}{P} - 1 \}$ (leaves of the parent tree).
			Then delete $OTPs_{LL}$ and $k$.
			Then compute $\mathcal{R} \leftarrow reduceMT(hOTPs)$ by \autoref{alg:MT-reduce-root}, the cached sublayer $cache$ of the first subtree and the proof $\pi_{sr}$ of that subtree's root hash $\mathcal{R}^s$ against $\mathcal{R}$.	
			Then create $tx_{constructor}(\mathcal{R}, 
			~cache, ~\pi_{sr})$ and send it to $\mathbb{W}$.
			Upon receiving $\{tx_{constructor}(\mathcal{R}, ~cache, ~\pi_{sr}, ~PK_\mathbb{U})\}$ from $\mathbb{W}$, forward it to $\mathbb{S}$.
			Upon receiving the event $ContrDeployed(\mathbb{S}^{ID})$ from $\mathbb{S}$, update UI and inform $\mathbb{U}$ about the deployment and display $\mathbb{S}^{ID}$.						
		}
		\item{\textbf{User $\mathbb{U}$:}			
			Once $k$ is generated by $\mathbb{A}$, transfer $k$ from $\mathbb{A}$ to $\mathbb{C}$ in an air-gapped manner.		
			Once $\mathbb{C}$ displays $\mathbb{S}^{ID}$, record $\mathbb{S}^{ID}$ as a public reference to $\mathbb{S}$. 
		}
		\item{\textbf{Private Key Wallet $\mathbb{W}$:}
			Generate private/public key-pair $SK_{\mathbb{U}}$, $PK_{\mathbb{U}} \leftarrow \Sigma.KeyGen()$.
			Upon receiving $tx_{constructor}(\mathcal{R}, ~cache)$ from $\mathbb{C}$, add $PK_{\mathbb{U}}$ to this transaction and send it to $\mathbb{C}$.
		}	
		\item {\textbf{Smart Contract $\mathbb{S}$:}
			Upon receiving $\{tx_{constructor}$ $(\mathcal{R}, ~cache, ~\pi_{sr}, ~PK_{\mathbb{U}})\}$ from $\mathbb{C}$, deploy the code of $\mathbb{S}$  (i.e., \autoref{alg:wallet-overview} enriched by storing of $cache$) on the blockchain, assigning $\mathbb{S}^{ID}$ to $\mathbb{S}$. 
			During the deployment, store $\mathcal{R}$, $PK_\mathbb{U}$, $cache$, 
			and adjust $nextOpID \leftarrow 0$.
			Next, compute root hash from $currentSubLayer$ and verify its consistency against $\mathcal{R}$ using $\pi_{sr}$.			
			Finally, send event $ContrDeployed(\mathbb{S}^{ID})$ to $\mathbb{C}$.
		}			
	\end{asparaitem}
	
\end{oframed}

\begin{oframed}
	\begin{center}
		\vspace{-0.2cm}
		Bootstrapping -- protocol $\Pi_{B}^\mathcal{I}$
		\\(for an insecure environment)
	\end{center}

	\begin{asparaitem}
		\item{\textbf{Authenticator $\mathbb{A}$:}
			Generate $k \leftarrow random()$. 
			Once $\mathbb{U}$ enters $N$, $N_S$, and $P$ to $\mathbb{A}$, compute $OTPs_{LL} \leftarrow F_k(\eta * \frac{N}{P} + i), ~i \in \{0, \ldots, \frac{N}{P} - 1\}$.			
			Then compute $hOTPs \leftarrow h^{P}_{\mathcal{D}}(OTPs_{LL}[i]), ~i \in \{0, \ldots,  \frac{N}{P} - 1 \}$ 
			and export them to microSD card.
			Then compute  $\mathcal{R} \leftarrow reduceMT(hOTPs)$ by \autoref{alg:MT-reduce-root} and display it to $\mathbb{U}$. 			
		}
		\item{\textbf{Client $\mathbb{C}$:}
			Upon delivering $hOTPs$ by $\mathbb{U}$ to $\mathbb{C}$, store them in the local storage.
			Then compute $root \leftarrow reduceMT(hOTPs)$ by \autoref{alg:MT-reduce-root}, the cached sublayer $cache$ of the first subtree and the proof $\pi_{sr}$ of that subtree's root hash $\mathcal{R}^s$. 
			Then create $tx_{constructor}(\mathcal{R}, ~cache, ~\pi_{sr})$ and send it to $\mathbb{W}$.
			Upon receiving $\{tx_{constructor}(\mathcal{R}, ~cache, ~\pi_{sr}, ~PK_\mathbb{U})\}$ from $\mathbb{W}$, forward it to $\mathbb{S}$.
			Upon receiving event $ContrDeployed(\mathbb{S}^{ID})$ from $\mathbb{S}$, inform $\mathbb{U}$ in UI.						
		}
		\item{\textbf{User $\mathbb{U}$:}
			Enter $N$, $N_S$, and $P$ to $\mathbb{A}$ and $\mathbb{C}$.
			Upon $hOTPs$ are exported by $\mathbb{A}$ to microSD card, transfer them to $\mathbb{C}$.						
			Upon $\mathcal{R}'$ of $\{tx_{constructor}(\mathcal{R}', ~cache, ~\pi_{sr})\}$ is displayed at $\mathbb{W}$, verify whether $\mathcal{R} = \mathcal{R}'$ by reading displays of $\mathbb{W}$ and $\mathbb{A}$. 
			In the positive case, proceed with the deployment by pressing a hardware button of $\mathbb{W}$.
			Once $\mathbb{W}$ displays $\mathbb{S}^{ID}$, record it as a public reference. 
		}
		\item{\textbf{Private Key Wallet $\mathbb{W}$:}
			Generate private/public key-pair $SK_{\mathbb{U}}$, $PK_{\mathbb{U}} \leftarrow \Sigma.KeyGen()$.
			Upon receiving $tx_{constructor}(\mathcal{R}', ~cache, ~\pi_{sr})$ from $\mathbb{C}$, display $\mathcal{R}'$ and $\mathbb{S}^{ID} \leftarrow h(PK_\mathbb{U} ~\|~ \mathcal{R}')$ to $\mathbb{U}$.
			Upon confirmation by $\mathbb{U}$, add $PK_{\mathbb{U}}$ to this transaction and send it to $\mathbb{C}$.
		}	
		\item {\textbf{Smart Contract $\mathbb{S}$:}
			The same as in $\Pi_{B}^{\mathcal{S}}$. 
			The only difference in contrast to $\Pi_{B}^\mathcal{S}$ is the requirement of a deterministic computation of $\mathbb{S}^{ID}$ by a blockchain platform using both $PK_\mathbb{U}$ and $\mathcal{R}$. 
			Hence $\mathbb{S}^{ID}$ can be computed by $\mathbb{W}$ and $\mathbb{S}$ independently.
		}	
		
	\end{asparaitem}
	
\end{oframed}

\begin{oframed}
	\begin{center}
		\vspace{-0.2cm}
		Operation execution -- protocol $\Pi_{O}$
	\end{center}

	\begin{asparaitem}
		\item{\textbf{Authenticator $\mathbb{A}$:}
			Upon receiving $opID$ from $\mathbb{U}$, compute $OTP_{opID} \leftarrow h^{\alpha(opID)}_{\mathcal{D}} ( F_k ( \beta(opID) ) ),$ where $\alpha(opID)$ and $\beta(opID)$ are computed by \autoref{eqn:alpha-beta-final}.
			Then display $OTP_{opID}$ to $\mathbb{U}$.
		}
		\item{\textbf{Client $\mathbb{C}$:}
			Once $args$ are entered by $\mathbb{U}$ into $\mathbb{C}$, construct $tx_{initOp}(args)$ and send it to $\mathbb{W}$.
			Upon receiving $\{tx_{initOp}(args)\}_{\mathbb{U}}$ from $\mathbb{W}$, forward it to $\mathbb{S}$.
			Upon receiving event $InitOpEvent(opID)$ from $\mathbb{S}$, update UI and inform $\mathbb{U}$ about initialization of $O_{opID}$.
			Upon entering $OTP_{opID}$ by $\mathbb{U}$, create proof $\pi_{opID}$ from the local storage.
			Then create $tx_{confirmOp}(OTP_{opID}, \pi_{opID}, opID)$ and send it to $\mathbb{S}$.
			Upon receiving event $ConfirmOpEvent(opID)$ from $\mathbb{S}$, update UI and inform $\mathbb{U}$.						 
		}
		\item{\textbf{User $\mathbb{U}$:}
			Enter $args$ of an operation into $\mathbb{C}$.
			Upon $args'$ of $tx_{initOp}(args')$ are displayed at $\mathbb{W}$, verify whether $args = args'$ by reading display of $\mathbb{W}$ and UI of $\mathbb{C}$.
			In the positive case, confirm signing of transaction by a hardware button of $\mathbb{W}$.
			Once $\mathbb{C}$ informs about initialized $O_{opID}$, enter $opID$ into $\mathbb{A}$.
			Once $\mathbb{A}$ displays $OTP_{opID}$, transfer $OTP_{opID}$ to $\mathbb{C}$ in an air-gapped manner.			
		}
		\item{\textbf{Private Key Wallet $\mathbb{W}$:}
			Upon receiving $tx_{initOp}(args')$ from $\mathbb{C}$, display $args'$ to $\mathbb{U}$.
			Upon confirmation of $args'$ by $\mathbb{U}$, sign $tx_{initOp}(args')$ by $\Sigma.Sign(tx, ~SK_\mathbb{U})$ and send it to $\mathbb{C}$.
		}	
		\item {\textbf{Smart Contract $\mathbb{S}$:}
			Upon receiving $\{tx_{initOp}(args)\}_{\mathbb{U}}$ from $\mathbb{C}$, verify signature $tx.\sigma$ by $\Sigma.Verify(tx.\sigma, ~PK_{\mathbb{U}})$.
			Then create a new operation $O_{opID}$ with $opID \leftarrow nextOpID$ using $args$ and increment  $nextOpID$. 
			Then send $InitOpEvent(opID)$ to $\mathbb{C}$.
			Upon receiving $tx_{confirmOp}(OTP_{opID}, \pi_{opID}, opID)$ from $\mathbb{C}$, verify $O_{opID}.pending = true$. 
			Then verify correctness of $OTP_{opID}$ by checking $current\-Sub\-Layer[(opID ~\%~ (N_S ~/~ P)) ~/~ 2^{H_S - L_S}]$ $=$ $derive\-Node\-InCache(OTP_{opID}, \pi_{opID}, opID)$ from \autoref{alg:MT-deriveNodeInCache-hash-chains} (or alternatively $\mathcal{R} = derive\-RootHash(OTP_{opID}, \pi_{opID}, opID)$ from \autoref{alg:MT-deriveRH-hash-chains} for the version without subtrees).
			Then execute $O_{opID}$ and set $O_{opID}.pending \leftarrow false$.
			Finally, send $ConfirmOpEvent(opID)$ to $\mathbb{C}$.
		}			
	\end{asparaitem}
	
\end{oframed}

\begin{oframed}
	\begin{center}
		\vspace{-0.2cm}		
		Introduction of a new parent tree -- protocol $\Pi_{NR}^\mathcal{S}$\\
		(for a secure environment)		
	\end{center}
	
	\begin{asparaitem}
		\item{\textbf{Authenticator $\mathbb{A}$:}			
			Once $\mathbb{U}$ enters $opID$ into $\mathbb{A}$, check whether $opID ~\%~ N = N-1$, and if so, notify $\mathbb{U}$ that a new parent tree is being introduced and display $k$ to $\mathbb{U}$.		
			Then compute $OTP_{opID} \leftarrow h^{\alpha(opID)} ( F_k ( \beta(opID) ) )$.
			Next, compute $OTPs_{LL} \leftarrow F_k(\eta \frac{N}{P} + i), ~i \in \{0, \ldots, \frac{N}{P} - 1\}$, where $\eta \leftarrow \eta + 1$.
			Then compute $\mathcal{R}^{new} \leftarrow aggregateOTPs(OTPs_{LL})$ by \autoref{alg:MT-reduce-root} and $hRootAndOTP \leftarrow  h(\mathcal{R}^{new} ~\|~ OTP_{opID})$.	
			Finally, show $\mathcal{R}^{new}$ and $hRootAndOTP$ to $\mathbb{U}$. 
			
		}
		\item{\textbf{Client $\mathbb{C}$:}		
			$\bullet$[\textit{Stage I}] $\mathbb{C}$ notifies $\mathbb{U}$ that a new parent tree needs to be introduced and displays $opID = N-1 + \eta N$, $\eta \in \{0, 1, \ldots \}$.
			Once $\mathbb{U}$ enters $k$ into $\mathbb{C}$,  compute $OTP_{N-1} \leftarrow h^{\alpha(opID)} ( F_k ( \beta(opID) ) ),$ where $\alpha(opID)$ and $\beta(opID)$ are computed by \autoref{eqn:alpha-beta-final}.
			Then create proof $\pi_{opID}$ from the local storage.  			
			Then compute $OTPs_{LL} \leftarrow F_k(\eta \frac{N}{P} + i), ~i \in \{0, \ldots, \frac{N}{P} -1\}$, where $\eta \leftarrow \eta + 1$. 
			Then compute and store $hOTPs \leftarrow h^{P}_{\mathcal{D}}(OTPs_{LL}[i]), ~i \in \{0, \ldots,  \frac{N}{P} - 1 \}$.
			Then delete $OTPs_{LL}$ and $k$.
			Then compute $\mathcal{R}^{new} \leftarrow reduceMT(hOTPs)$ by \autoref{alg:MT-reduce-root}.				 			
			Then compute $hRootAndOTP \leftarrow h(\mathcal{R}^{new} ~\|~ OTP_{opID})$, construct $tx_{1\_newRootHash}(hRootAndOTP)$, and send it to $\mathbb{W}$.
			Upon receiving $\{tx_{1\_newRootHash}(hRootAndOTP)\}_{\mathbb{U}}$ from $\mathbb{W}$, forward it to $\mathbb{S}$.
			$\bullet$[\textit{Stage II}] Upon $newRootHash1(hRootAndOTP)$ event is received from $\mathbb{S}$, construct $tx_{2\_newRootHash}(\mathcal{R}^{new})$ and send it to $\mathbb{W}$.
			Once $\{tx_{2\_newRootHash}(\mathcal{R}^{new})\}_{\mathbb{U}}$ is received from $\mathbb{W}$, forward it to $\mathbb{S}$. 
			$\bullet$[\textit{Stage III}] Upon receiving the event $newRootHash2(\mathcal{R}^{new})$ from $\mathbb{S}$, compute the cached sublayer $cs$ of the first subtree in the new parent tree and the proof $\pi_{sr}$ of the subtree's root hash $\mathcal{R}^s$. 
			Then construct $tx_{3\_newRootHash}(OTP_{opID}, ~\pi_{opID}, ~cs, ~\pi_{sr})$ and send it to $\mathbb{S}$.
			Upon receiving event $newRootHash3(OTP_{opID})$ from $\mathbb{S}$, update UI and inform $\mathbb{U}$.
			
		}
		\item{\textbf{User $\mathbb{U}$:}
			Once $\mathbb{C}$ displays $opID = N-1 + \eta N$, $\eta \in \{0, 1, \ldots \}$ and informs $\mathbb{U}$ about the necessity of introducing a new parent tree, enter $opID$ into $\mathbb{A}$.
			Once $\mathbb{A}$ displays $k$ and shows a message that a new parent tree is being introduced, transfer $k$ from $\mathbb{A}$ to $\mathbb{C}$ in an air-gapped manner.
			Once $hRootAndOTP'$ of $\{tx_{1\_newRootHash}(hRootAndOTP')\}$ is displayed at $\mathbb{W}$, verify $hRootAndOTP' = hRootAndOTP$ by reading displays of $\mathbb{W}$ and $\mathbb{A}$. 
			Once $\mathcal{R}^{new'}$ of $\{tx_{2\_newRootHash}(\mathcal{R}^{new'})\}$ is displayed at $\mathbb{W}$, verify $\mathcal{R}^{new'} = \mathcal{R}^{new}$ by reading displays of $\mathbb{W}$ and $\mathbb{A}$. 
			If so, confirm signing by a hardware button of $\mathbb{W}$.
			
		}
		\item{\textbf{Private Key Wallet $\mathbb{W}$:}
			The same as in $\Pi_{O}$.
		}	
		\item {\textbf{Smart Contract $\mathbb{S}$:}
			$\bullet$[\textit{Stage I}] Upon receiving $\{tx_{1\_newRootHash}(hRootAndOTP)\}_{\mathbb{U}}$ from $\mathbb{C}$, verify signature $tx.\sigma$ by $\Sigma.Verify(tx.\sigma, ~PK_{\mathbb{U}})$.
			Then verify  $nextOpID ~\%~ N = N - 1$; if so, append $hRootAndOTP$ into $L_1$.
			Then send event $newRootHash1(hRootAndOTP)$ to $\mathbb{C}$.
			$\bullet$[\textit{Stage II}] Upon receiving $\{tx_{2\_newRootHash}(\mathcal{R}^{new})\}_{\mathbb{U}}$ from $\mathbb{C}$, verify signature $tx.\sigma$ by $\Sigma.Verify(tx.\sigma, ~PK_{\mathbb{U}})$.
			Then verify  $nextOpID ~\%~ N = N - 1$; if so, append $\mathcal{R}^{new}$ into $L_2$.
			Then send event $newRootHash2(\mathcal{R}^{new})$ to $\mathbb{C}$.
			$\bullet$[\textit{Stage~III}] Once $\{tx_{3\_newRootHash}(OTP_{opID}, ~\pi_{opID}, ~cs, \pi_{sr})\}$ is received from $\mathbb{C}$, verify  $nextOpID ~\%~ N = N - 1$.
			Then verify correctness of $OTP_{opID}$ by checking $currentSubLayer[(opID ~\%~ (N_S ~/~ P)) ~/~ 2^{H_S - L_S}] = deriveNodeInCache(OTP_{opID}, ~\pi_{opID}, ~opID)$ from \autoref{alg:MT-deriveNodeInCache-hash-chains} (or $\mathcal{R} = deriveRootHash(OTP_{opID},$ $ ~\pi_{opID}, ~opID)$ from \autoref{alg:MT-deriveRH-hash-chains} in the version without subtrees).
			Then locate the first entries of $L_1$ and $L_2$ that match the condition $h(L_2[i] ~\|~ OTP_{opID}) = L_1[j]$. 
			If matching entries are found, then set $\mathcal{R} \leftarrow L_2[i]$, increment $nextOpID$, adjust $currentSubLayer \leftarrow cs$ and verify its consistency
			 by $subtreeConsistency(\mathcal{R}^s, ~\pi_{sr}, ~\mathcal{R})$, where $\mathcal{R}^s \leftarrow reduceMT(currentSubLayer, currentSubLayer.len)$.
			Finally, clear the lists $L_1, L_2 \leftarrow [], []$.
			
		}	
		
	\end{asparaitem}
	
\end{oframed}

\begin{oframed}
	\begin{center}
		\vspace{-0.2cm}		
		Introduction of a new parent tree -- protocol $\Pi_{NR}^\mathcal{I}$\\
		(for an insecure environment)		
	\end{center}
	
	\begin{asparaitem}
		\item{\textbf{Authenticator $\mathbb{A}$:}			
			Once $\mathbb{U}$ enters $opID$ into $\mathbb{A}$, check whether $opID ~\%~ N = N-1$, and if so display $OTP_{opID} \leftarrow h^{\alpha(opID)} ( F_k ( \beta(opID)))$ and notify $\mathbb{U}$ that a new parent tree is being introduced.
			Then, compute $OTPs_{LL} \leftarrow F_k(\eta \frac{N}{P} + i), ~i \in \{0, \ldots, \frac{N}{P} - 1\}$, where $\eta \leftarrow \eta + 1$.
			Then compute $hOTPs \leftarrow h^{P}(OTPs_{LL}[i]), ~i \in \{0, \ldots,  \frac{N}{P} - 1 \}$  and export it to a microSD card.
			Finally, compute $\mathcal{R}^{new} \leftarrow reduceMT(hOTPs)$ by \autoref{alg:MT-reduce-root} and $hRootAndOTP \leftarrow h(\mathcal{R}^{new} ~\|~ OTP_{opID})$, and display both to $\mathbb{U}$. 					
		}
		\item{\textbf{Client $\mathbb{C}$:}		
			[\textit{Stage I}] $\mathbb{C}$ notifies $\mathbb{U}$ that a new parent tree needs to be introduced and displays $opID = N-1 + \eta N, ~\eta \in \{0,1, \ldots\}$.
			Upon entering $OTP_{opID}$ by $\mathbb{U}$, create proof $\pi_{opID}$ from the local storage.
			Once leaves of the tree $hOTPs$ are delivered by $\mathbb{U}$ into $\mathbb{C}$, store $hOTPs$ in the local storage.
			Then compute $\mathcal{R}^{new} \leftarrow reduceMT(hOTPs)$ by \autoref{alg:MT-reduce-root}. 			
			Then compute $hRootAndOTP \leftarrow h(\mathcal{R}^{new} ~\|~ OTP_{opID})$, construct $tx_{1\_newRootHash}(hRootAndOTP)$, and send it to $\mathbb{W}$.
			Upon receiving $\{tx_{1\_newRootHash}(hRootAndOTP)\}_{\mathbb{U}}$ from $\mathbb{W}$, forward it to $\mathbb{S}$.
			[\textit{Stages II}] and [\textit{Stage III}] are the same as in $\Pi_{NR}^{\mathcal{S}}$
			
		}
		\item{\textbf{User $\mathbb{U}$:}
			Once $\mathbb{C}$ displays $opID = N-1 +\eta N, ~\eta \in \{0,1, \ldots\}$ and informs $\mathbb{U}$ about necessity of introducing a new parent tree, enter $opID$ into $\mathbb{A}$.
			Once $\mathbb{A}$ displays $OTP_{opID}$ and notify $\mathbb{U}$ that the new parent tree is being introduced, transfer $OTP_{opID}$ to $\mathbb{C}$ in an air-gapped manner.
			Once $hOTPs$ are exported by $\mathbb{A}$ to microSD card, transfer them to $\mathbb{C}$.						
			Once $hRootAndOTP'$ of $\{tx_{1\_newRootHash}(hRootAndOTP')\}$ is displayed at $\mathbb{W}$, verify $hRootAndOTP' = hRootAndOTP$ by reading displays of $\mathbb{W}$ and $\mathbb{A}$; in the positive case, confirm signing of transaction within $\mathbb{W}$ by a hardware button.								
			Once $\mathcal{R}^{new'}$ of $\{tx_{2\_newRootHash}(\mathcal{R}^{new'})\}$ is displayed at $\mathbb{W}$, verify $\mathcal{R}^{new'} = \mathcal{R}^{new}$ by reading displays of $\mathbb{W}$ and $\mathbb{A}$;
			in the positive case, confirm signing of transaction within $\mathbb{W}$ by a hardware button.			
		}
		\item{\textbf{Private Key Wallet $\mathbb{W}$:}
			The same as in $\Pi_{O}$.
		}	
		\item {\textbf{Smart Contract $\mathbb{S}$:}		
			The same as in $\Pi_{NR_S}$.
		}	
		
	\end{asparaitem}
	
\end{oframed}

\begin{oframed}
	\vspace{-0.2cm}
	\begin{center}
		Introduction of the next subtree -- protocol $\Pi_{ST}$
	\end{center}
	
	\begin{asparaitem}
		\item{\textbf{Authenticator $\mathbb{A}$:}			
			The same as in $\Pi_{O}$, while in addition, $\mathbb{A}$ displays a message that the next tree is being introduced.
		}
		\item{\textbf{Client $\mathbb{C}$:}		
			$\mathbb{C}$ notifies $\mathbb{U}$ that a new subtree needs to be introduced and displays $opID = (N_S - 1) + \delta N_S, ~\delta \in \{0, ~\ldots,~ \frac{N}{N_S} - 2\}$.
			Upon entering $OTP_{opID}$ by $\mathbb{U}$, create the proof $\pi_{opID}$ from the local storage.
			Then compute $nextSubLayer$ (i.e., the cached sublayer of the next subtree)  and $\pi_{sr}$ (i.e., the proof of the next subtree's root) from $\mathbb{C}$'s storage.
			Next construct $tx_{nextSubtree}(nextSubLayer, ~OTP_{opID}, ~\pi_{opID}, ~\pi_{sr})$ and send it to $\mathbb{S}$.
			Upon receiving event $newSubtree(opID)$ from $\mathbb{S}$, update UI and inform~$\mathbb{U}$.
		}
		\item{\textbf{User $\mathbb{U}$:}
			Once $\mathbb{C}$ displays $opID = (N_S - 1) + \delta N_S, ~\delta \in \{0, ~\ldots,~ \frac{N}{N_S} - 2\}$ and informs $\mathbb{U}$ about necessity of introducing the next  subtree, enter $opID$ into $\mathbb{A}$.
			Once $\mathbb{A}$ displays $OTP_{opID}$ and confirming that the next tree is being introduced, transfer it to $\mathbb{C}$ in an air-gapped manner.

		}
		\item{\textbf{Private Key Wallet $\mathbb{W}$:}
			No interaction required.
		}	
		\item {\textbf{Smart Contract $\mathbb{S}$:}		
			Upon receiving $tx_{nextSubtree}(nextSubLayer, ~OTP_{opID}, ~\pi_{opID}, ~\pi_{sr})$ from $\mathbb{C}$, verify $nextOpID ~\%~N \neq N - 1 \wedge nextOpID ~\%~N_S = N_S - 1 \wedge currentSubLayer.len = nextSubLayer.len$.
			Then verify correctness of $OTP_{opID}$ by checking $cache[(opID ~\%~ (N_S ~/~ P)) ~/~ 2^{H_S - L_S}] = deriveNodeInCache(OTP_{opID}, ~\pi_{opID}, ~opID)$ from \autoref{alg:MT-deriveNodeInCache-hash-chains}.
			Next, update the current cached sublayer $currentSubLayer \leftarrow nextSubLayer$ and check its consistency against $\mathcal{R}$ by a function $subtreeConsistency(\mathcal{R}^s, ~\pi_{sr}, ~\mathcal{R})$.
			Note that this function requires already computed $\mathcal{R}$ of the next subtree using \autoref{alg:MT-reduce-root}: $\mathcal{R}^s \leftarrow reduceMT(currentSubLayer, ~currentSubLayer.len)$ and its proof $\pi_{sr}$.
			Finally, increment $nextOpID$ and send event $newSubtree(opID)$ to $\mathbb{C}$. 			
		}	
		
	\end{asparaitem}
	
\end{oframed}

\begin{figure}[h]
	\begin{center}
		\includegraphics[width=0.5\textwidth]{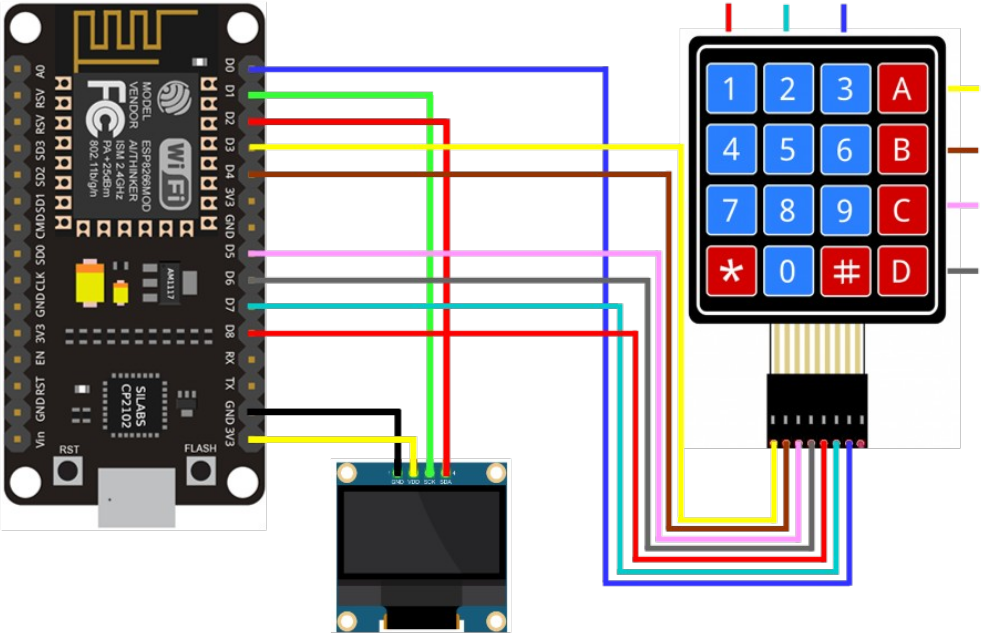} 
		\vspace{-0.1cm}
		\caption{The scheme of a proof-of-concept hardware implementation of the authenticator.}
		\label{fig:gw-auth}
	\end{center}	
\end{figure}

\newpage
\subsection{Hardware Implementation of $\mathbb{A}$}\label{appendix:NodeMcu-schema}
For demonstration purposes, we constructed a proof-of-concept hardware implementation of the authenticator of SmartOTPs using cheap hardware and C language.
In detail, we selected NodeMCU with ESP8266 MCU that costs around $\$$2.
Next, we selected a 0.96" OLED display with resolution of 128x64 that was connected to MCU by 4-wire SPI interface (the cost of such a display falls below $\$$2). 
To control the display, we used Adafruit\_SSD1306 a Adafruit\_GFX libraries.
Finally, we used a simple 4x4 keyboard with 3+4 wires addressing a combination of 3 columns and 4 rows (the cost of the keyboard is around $\$$0.5).
To interact with the keyboard, we utilized Keypad library of Adruino that is built for matrix style keyboards.
Further, we used software version of hash function, available from \url{https://github.com/ethereum/ethash}.
The scheme of our hardware implementation is depicted in \autoref{fig:gw-auth} and
the source with a demonstration video is provided at \url{https://www.dropbox.com/sh/gmcz8zt12j7omsf/AADR4LHDOhSlwANnI707gkMda?dl=0}.
\vspace{1cm}

%\subsection{Overview of Wallets}\label{appendix-soa}
%

\end{document}